\documentclass[pre,singlecolumn,noshowpacs,preprintnumbers,amsmath,amssymb,nofootinbib,]{revtex4-2}
\usepackage{bbm}
\usepackage{lipsum}
\usepackage{mathrsfs}
\usepackage{graphicx}
\usepackage{dcolumn}
\usepackage{bm}
\usepackage{color}
\usepackage{comment}
\usepackage{hyperref}
\usepackage{times}
\usepackage{ulem}
\usepackage{ascmac}
\usepackage{footnote}

\newcommand{\tb}{\textbf}

\newcommand{\ra}{\rangle}
\newcommand{\la}{\langle}

\newcommand{\eps}{\epsilon}

\newcommand{\ID}{\mathfrak{i}}

\newcommand{\dd}{{\rm d}}

\newcommand{\rmE}{{\mathrm{E}}}

\newcommand{\Fourier}{\mathcal{F}}
\newcommand{\Laplace}{\mathcal{L}}
\newcommand{\ZTrans}{\mathcal{Z}}
\newcommand{\SetTrader}{\bm{\Omega}}
\newcommand{\Dt}{\Delta t}
\newcommand{\ini}{{\rm ini}}
\newcommand{\fin}{{\rm fin}}
\newcommand{\pd}{\partial}

\usepackage{tcolorbox} 
\tcbuselibrary{skins,breakable,theorems}
\newenvironment{box_summary}[1]{
    \begin{tcolorbox}[	colframe = blue!20, colback = blue!4, 
												coltitle=black, fonttitle=\bfseries, title = {#1}, breakable = false ]
}{
    \end{tcolorbox}
}

\begin{document}
\title{
		Exactly solvable model for the diffusive price-dynamics paradox \\ under long-range correlated market-order flow
}
\author{Yuki Sato}
\affiliation{Institute of Systems and Information Engineering, University of Tsukuba, Tennodai, Tsukuba, Ibaraki 305-8573, Japan
}
\email{sato.yuki.gb@u.tsukuba.ac.jp}

\author{Shunta Fujiwara}
\author{Kiyoshi Kanazawa}

\affiliation{Department of Physics, Graduate School of Science, Kyoto University, Kyoto 606-8502, Japan}
\date{\today}


\begin{abstract}
    We develop an exactly solvable nonlinear time-series model by incorporating the square-root price-impact law into the Lillo--Mike--Farmer (LMF) model to resolve the diffusive price-dynamics paradox under predictable market-order flow. In financial market microstructure, it is well established that the price dynamics are approximately described by Brownian motion at long times. However, it is also well-known that market-order flow is clearly predictable due to long-range correlations, as mathematically formulated by the LMF model. Since market orders have a positive price impact in general, predictable market-order flow seems to contradict Brownian price dynamics. In this work, we resolve this diffusive price-dynamics paradox by developing nonlinear time-series models that generalize the LMF model based on the square-root price-impact law. Our time-series models can be mathematically mapped onto the L\'evy-walk framework---an exactly solvable class of non-Markovian stochastic processes developed in statistical physics. We prove that the price dynamics are diffusive at long times under the square-root law even under predictable market-order flow. Our work highlights the crucial practical importance of the square-root law in understanding the microstructural foundation of the Efficient Market Hypothesis.
\end{abstract}

\maketitle

\section{Introduction}
	In financial economics, it is well established that predicting future stock prices is extremely difficult, and stock-price dynamics are approximately described by Brownian motion. Most modern financial theories---including option pricing and credit risk models---are built upon this assumption, and this empirical fact is formalized as the efficient market hypothesis (EMH), which serves as a cornerstone of modern finance.
	
		While the EMH plays a key role in modern finance theories, empirical studies have reported macroscopic phenomena that appear to contradict this picture~\cite{LMF_PRE2005,BouchaudText}. For example, the long-range correlation (LRC) of the buy-sell market order flow is an enigmatic phenomenon because it implies that the market-order flow is predictable for a long time. In financial markets, once a buy (sell) market order is observed, subsequent market orders are more likely to be of the same sign. This statistical regularity is known as LRC and is quantitatively characterized by the power-law decay of the autocorrelation function (ACF):
	\begin{align}
		C(\tau):= \rmE[\eps(t)\eps(t+\tau)] \propto \tau^{-\gamma},\>\>\> 0<\gamma<1.
        \label{eq:intro:LRC}
	\end{align}
	where $\gamma$ characterizes the LRC, and the order sign $\eps(t)=+1$ ($\eps(t)=-1$) denotes a buy (sell) market order. This long-memory phenomenon originates from the order-splitting behavior of institutional investors: in other words, institutional investors occasionally have a large metaorder and split it into a long sequence of child orders to minimize their price impact~\cite{LMF_PRE2005,SatoPRL2023,SatoPRR2023}. Since all the child orders have the same order sign, the market-order sign has persistent predictability as formulated by the LRC~\eqref{eq:intro:LRC}. This microscopic hypothesis was mathematically formalized as the Lillo--Mike--Farmer (LMF) model~\cite{LMF_PRE2005} and has been quantitatively verified in our recent Letter by analyzing a microscopic dataset on the Tokyo Stock Exchange (TSE)~\cite{SatoPRL2023,SatoPRR2023}.
    
    Given this empirical law, the standard price formation theory based on the EMH does not seem straightforwardly consistent with the LRC of the market-order flow, particularly when one assumes a linear price response to the order flow. To clarify this point, let us formulate the price impact as a function of the transaction volume. The price impact is defined as the average price change following market orders with cumulative volume $Q$, called metaorder size in the literature: 
    \begin{equation}
        I(Q) := \rmE[\eps \Delta p\mid Q]
    \end{equation}
    with the price change $\Delta p:=p(t_{\fin})-p(t_{\ini})$, where $t_{\ini}$ and $t_{\fin}$ are the initial and final times of the metaorder execution. Typical economic theories predict the linear response relation, such that 
    \begin{equation}
        I(Q) \approx \lambda Q,
    \end{equation}
    where the coefficient $\lambda$ is called Kyle's lambda, named after Kyle's pioneering work~\cite{Kyle1985}. If the price impact has a linear permanent component, called the permanent impact, price dynamics of linear price-impact models exhibit superdiffusion under the LRC, which is clearly inconsistent with the empirical Brownian motion. Thus, the empirical diffusive behavior of price dynamics is enigmatic in the presence of strongly predictable market-order flow, as formulated by the LRC~\eqref{eq:intro:LRC}. In this work, we refer to this apparent contradiction as the {\it diffusive price-dynamics paradox}.

    Importantly, detailed empirical analyses in the econophysics literature have demonstrated that the linear price-impact law is not empirically accurate for large $Q$; instead, the square-root law holds for various asset classes: 
	\begin{align}
		I(Q)\propto Q^{\delta},\>\>\>\delta\simeq \frac{1}{2}.
        \label{eq:intro-SRL}
	\end{align}
    This empirical law was initially found by hedge-fund groups, and some econophysics groups have claimed its universality through their investment experiments~\cite{BouchaudText}. Indeed, a recent comprehensive survey of the TSE conclusively supported such a universality argument quantitatively: the exponents $\delta$ were equal to one-half within statistical errors for all liquid stocks on the TSE~\cite{SatoPRL2025}. Despite its surprising robustness, its microscopic origin remains a mystery, making the square-root law one of the enigmatic empirical laws in financial economics.
    
	In this paper, we claim that the square-root law is the key to resolving this diffusive price-dynamics paradox, namely, the coexistence of diffusive price dynamics with the predictable market-order flow generated by the LMF mechanism. We propose an exactly solvable time-series model by incorporating the square-root law into the LMF model: the price impact obeys the nonlinear relation $I(Q) \propto Q^{\delta}$ under the predictable market-order flow characterized by $C(\tau)\propto \tau^{-\gamma}$, with $\delta \in (0,1)$ and $\gamma \in (0,1)$. For this model, we prove that $\delta\leq 1/2$ is the necessary and sufficient condition for the price dynamics to be diffusive for any $\gamma \in (0,1)$. Thus, the square-root law $\delta=1/2$ guarantees diffusive price dynamics even in the presence of LRC. We also propose several extensions of our nonlinear price-impact model, incorporating elements such as inter-metaorder resting periods and post-metaorder impact decay, to demonstrate the robustness of the diffusive behavior even under LRC. We further show that our minimal models are consistent with various empirical laws, such as the inverse-cubic law (ICL) and the volatility clustering, even though we just incorporate the square-root law into the LMF model. Our result implies that the square-root law suppresses the price impact of institutional investors, thereby leading to normal diffusion even under large-scale metaorder splitting, and that the square-root law plays a major role in reconciling predictable market-order flow with the EMH.

	This paper is organized as follows. Section~\ref{sec:notation} describes our mathematical notation and reviews the original LMF model. This section clarifies what is missing in the original LMF and what we incorporated into the LMF model as a minimal generalization. The main part starts from Sec.~\ref{sec:discrete-model}. Section~\ref{sec:discrete-model} introduces a discrete-time time-series model based on the LMF model and the square-root law. We show that price dynamics are diffusive even though market-order flow exhibits long memory if the price impact is sufficiently concave. We will consider a continuous-time version of our generalized LMF time-series model in Sec.~\ref{sec:continuous-time-model}. The advantage of the continuous-time model is that the pre-existing theory for the L\'evy-walk model is readily applicable and, thus, the formulation and the corresponding results can be easily understood. Section~\ref{sec:further-generalization} shows further generalizations of our model, where the consistency with the ICL and the volatility clustering are discussed. We conclude this paper with several remarks in Sec.~\ref{sec:conclusion}. Four appendices supplement technical parts of the main results.

\section{Notation and literature review}\label{sec:notation}
    In this section, we describe our mathematical notation and review the original LMF model to clarify the motivation of our study.

    \subsection{Notation for probability theory}
        Let us first introduce some mathematical notation for probability theory. We denote the probability density function (PDF) of a time-independent random variable $X$ by $P(X)$. For a time-dependent process $X(t)$, its PDF is written as $P(X,t)$ and its complementary cumulative distribution function (CCDF) is written as 
        \begin{equation}
            P_{\geq}(X):=\int_X^\infty \dd X'P(X').
        \end{equation}
        
        The ensemble average of a random variable $X$ is defined as 
        \begin{equation}
            \rmE[X] = \int\dd X P(X) X,\>\>\>\rmE[X] = \sum_{X} P(X)X
        \end{equation}
        for continuous and discrete variables, respectively. The conditional PDF is defined as $P(A\mid B) = P(A,B)/P(B)$, and the conditional expectation is defined as
        \begin{equation}
            \rmE[A\mid B] = \int \dd{A} P(A\mid B) A,\>\>\> \rmE[A\mid B] = \sum_A P(A\mid B) A
        \end{equation}
        for continuous and discrete variables, respectively. 
        
        In addition, the indicator function $\mathbb{I}_A$ is defined for a mathematical statement $A$, such that $\mathbb{I}_A=1$ if $A$ is true; otherwise $\mathbb{I}_A=0$ (e.g., $\mathbb{I}_{x\geq 1}=1$ for $x=2$ and $\mathbb{I}_{y\geq 1}=0$ for $y=0$). The indicator function is formally related to Dirac's $\delta$ function $\delta(x)=\lim_{\eps \to 0}\eps^{-1}\mathbb{I}_{|x|\leq \eps/2}$. The PDF is rewritten as $P(X=x)=\la\delta(X-x)\ra$ for a continuous stochastic variable $X$. 

    \subsection{Notation for integral transformations}
        We introduce our notation for integral transforms. The Fourier and Laplace transforms of a function $f(X,t)$ with respect to the continuous variable $X$ are written as  
        \begin{align}
            \Fourier_{X\to k}[f(X,t)]:= \int^{\infty}_{-\infty} \dd X e^{-ikX} f(X,t),\>\>\>
            \Laplace_{t\to s}[f(X,t)] := \int_{0}^{\infty} \dd t\, e^{-st} f(X,t).
        \end{align}
        For example, the characteristic function of a process $X(t)$ with PDF $P(X,t)$ is given by $\Fourier_{X\to k}[P(X,t)]$. We also introduce the $z$-transform of a function $f(X,t)$ with respect to the discrete time variable $t$ as 
        \begin{align}
            \ZTrans_{t\to z}[f(X,t)]:= \sum_{t=0}^{\infty} f(X,t)\, \frac{z^t}{t!}.
        \end{align}
        
    \subsection{Notation in continuous-time random-walk theory}
        In continuous-time random-walk (CTRW) theory, the Fourier transform is always applied to spatial variables (such as a position $X$ and price $p$), such that $\Fourier_{X\to k}[f(X,t)]$. Also, the Laplace transform is always applied to continuous time $t$, such that $\Laplace_{t\to s}[f(X,t)]$. Therefore, it is customary to use the abbreviated notation for the Fourier and Laplace transforms, such that 
        \begin{align}
            f(k,t) := \Fourier_{X\to k}[f(X,t)], \quad 
            f(X,s) := \Laplace_{t\to s}[f(X,t)], \quad 
            f(k,s) := \left(\Fourier_{X\to k}\circ \Laplace_{t\to s}\right)[f(X,t)],
        \end{align}
        where the arguments $k$ and $s$ signify the Fourier and Laplace representations, respectively. Similarly, the $z$-transform is applied to discrete time $t$, and we introduce an abbreviated notation, such that 
        \begin{equation}
            f(X,z) := \ZTrans_{t\to z}[f(X,t)], \quad 
            f(k,z) := \left(\Fourier_{X\to k}\circ \ZTrans_{t\to z}\right)[f(X,t)],
        \end{equation}
        where the arguments $k$ and $z$ signify the Fourier and $z$ representations, respectively. 

    \subsection{Review of the original LMF model}
        In this subsection, we will review the original LMF model~\cite{LMF_PRE2005}. Whereas a price-generating mechanism is missing in the original LMF model, we will add minimal price-generating dynamics to the LMF model in the main part from Sec.~\ref{sec:discrete-model} to Sec.~\ref{sec:further-generalization}.
        
        \subsubsection{State-variable space}
            The LMF model is composed of several order-splitting traders. The total number of order splitters is denoted by a positive integer $M\geq 1$. The set of order-splitting traders is defined as $\SetTrader:=\{1,2,\dots,M\}$. In this paper, any quantity with a superscript in parentheses, such as $X^{(i)}$, denotes a state variable for an individual trader with trader ID $i \in \SetTrader$. Sets are denoted in {\it bold} font, while other variables are written in {\it italic} font.

            In this paper, we consider the simplest version of the LMF model: all traders have the same order-submission intensity at each time step $\lambda^{(i)} = \lambda=1/M$ for all $i \in \SetTrader$. Also, we assume that the market-order size of any child order is the unit volume, and that the metaorder size distribution is given by a power-law PDF: 
            \begin{equation}
                \psi_{m}(L) = \frac{1}{\zeta(\alpha+1)} L^{-\alpha-1}, \quad \alpha > 1
            \end{equation}
            for a positive integer $L$, where $\zeta(x):=\sum_{n=1}^\infty 1/n^{x}$ is the Riemann zeta function. Note that $\alpha>1$ is a necessary condition for the LMF model to have a stationary ACF. Also, $\alpha \in (1,2)$ is the typical value range observed in empirical analyses of the LRC.
            
            The state variables for the $i$-th order splitter are given by the metaorder sign $\eps^{(i)}(t) \in \{-1, +1\}$, the remaining metaorder volume $R^{(i)}(t)$, and the executed metaorder volume $Q^{(i)}(t)$, where $i \in \SetTrader$. The market-level order sign is characterized by $\eps(t)$. The state of the original LMF model is completely characterized by a point in the phase space 
            \begin{align}
                \Xi(t):=\left(\eps(t); \eps^{(1)}(t), R^{(1)}(t), Q^{(1)}(t);\dots;  \eps^{(M)}(t), R^{(M)}(t), Q^{(M)}(t)  \right).
            \end{align}
            Thus, the LMF model is formulated as a $(3M+1)$-dimensional Markov process.

      \subsubsection{Stochastic dynamics of the order flow}
            \begin{figure*}
                \includegraphics[width=140mm]{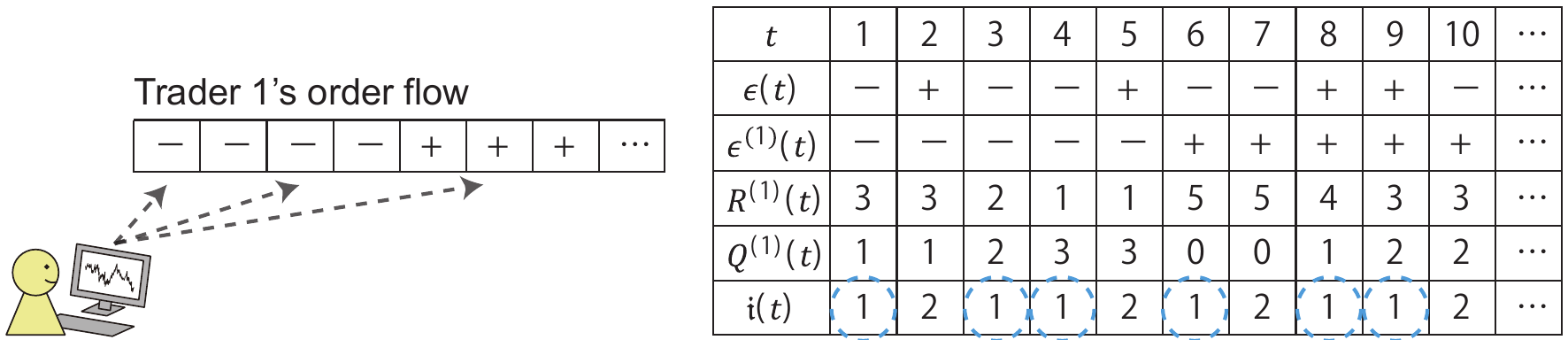}
                \caption{
                    Schematic of the Lillo--Mike--Farmer model. For each $i \in \SetTrader$, the internal state of trader $i$ is characterized by $(\eps^{(i)}, R^{(i)}, Q^{(i)})$, where $\eps^{(i)}$ is the metaorder sign, $R^{(i)}$ is the remaining metaorder volume, and $Q^{(i)}$ is the cumulative executed metaorder volume. In the coarse-grained description, the market-order sign $\eps(t)$ is determined by $\eps(t)=\eps^{(\ID(t))}(t)$, where $\ID(t)$ denotes the ID of the trader submitting the order at time $t$. Since the metaorder size $L$ obeys a power-law PDF, traders occasionally have large metaorders, which are split into long sequences of child orders. Because all child orders within the same metaorder have the same sign, the market-order sign $\eps(t)$ exhibits long memory.
                }\label{fig:schematic-LMF}
            \end{figure*}
      
            We next formulate the stochastic order-flow dynamics of the LMF model (see Fig.~\ref{fig:schematic-LMF}). Let us introduce a random variable $\ID(t)$, which represents the trader who submits a market order at time $t$. In the LMF model, each trader submits market orders with intensity $\Lambda$. In the discrete-time formulation with time step $\Dt$, the probability that trader $i$ submits an order at time $t$ is given by $P(\ID(t) = i) = \Lambda\Dt =: \lambda$ for all $i \in \SetTrader$. After a trader submits a market order, the executed volume $Q^{(\ID)}(t)$ is incremented and the remaining volume $R^{(\ID)}(t)$ is decremented by a unit volume, provided that $R^{(\ID)}(t) > 1$. When the metaorder is completely executed (i.e., $R^{(\ID)}(t+\Dt)=R^{(\ID)}(t)- 1=0$), the trader's state is reinitialized: a new sign $\eps^{(\ID)}$ is drawn uniformly from $\{-1,+1\}$, a new metaorder size $L$ is drawn from $\psi_{m}(L)$, and the executed volume $Q^{(\ID)}$ is reset to zero. More precisely, the stochastic dynamics is given by 
            \begin{box_summary}{Complete stochastic dynamics of the original LMF model}\vspace{-5mm}
            \begin{subequations}\label{eq:review-LMF-SDE}
                \begin{align}
                    R^{(i)}(t+\Dt) &= 
                    \begin{cases}
                        R^{(i)}(t) & \mbox{if $i \neq \ID(t+\Dt)$} \\
                        R^{(i)}(t) - 1& \mbox{if $i = \ID(t+\Dt)$ and $R^{(i)}(t)>1$} \\
                        L & \mbox{if $i= \ID(t+\Dt)$ and $R^{(i)}(t)=1$, $L$ obeys $\psi_{m}(L)$}
                    \end{cases}, \\	
                    Q^{(i)}(t+\Dt) &= 
                    \begin{cases}
                        Q^{(i)}(t) & \mbox{if $i \neq \ID(t+\Dt)$} \\
                        Q^{(i)}(t) + 1 & \mbox{if $i = \ID(t+\Dt)$ and $R^{(i)}(t)>1$} \\
                        0 & \mbox{if $i= \ID(t+\Dt)$ and $R^{(i)}(t)=1$}
                    \end{cases}, \\
                    \eps^{(i)}(t+\Dt) &= 
                        \begin{cases}
                        \eps^{(i)}(t) & \mbox{if $i \neq \ID(t+\Dt)$ or $R^{(i)}(t)>1$} \\
                            +1 & \mbox{with prob. $1/2$, if $i = \ID(t+\Dt)$ and $R^{(i)}(t)=1$} \\
                            -1 & \mbox{with prob. $1/2$, if $i = \ID(t+\Dt)$ and $R^{(i)}(t)=1$}
                    \end{cases}, \\
                    \eps(t+\Dt) &= \eps^{( \ID(t+\Dt))}(t+\Dt)
                \end{align}
            \end{subequations}
            \end{box_summary}\noindent
            where $\Dt$ denotes the unit time step. In the following, we set $\Dt=1$ for simplicity. 

        \subsubsection{Order-sign ACF}
            The ACF of the order sign $\eps(t)$ asymptotically decays according to a power law~\cite{LMF_PRE2005,GeneralizedLMF}: 
            \begin{box_summary}{LRC in the order-sign ACF for the LMF model}\vspace{-5mm}
                \begin{equation}
                    C(\tau) \propto \tau^{-\gamma},\>\>\>\gamma=\alpha-1 
                \end{equation}
            \end{box_summary}\noindent
            for $\tau\gg 1$. This behavior is consistent with the LRC in empirical analyses. This asymptotic decay can be proved by an exact solution~\cite{GeneralizedLMF}. This power-law relation was empirically confirmed in our recent works~\cite{SatoPRL2023,SatoPRR2023}.

        \subsubsection{Goal of this work}
            The above review clarifies that the original LMF model addresses only the generating mechanism of market-order flow and lacks a price-generating mechanism. In this work, we incorporate the nonlinear price impact of metaorders, i.e., the square-root law~\eqref{eq:intro-SRL}, and develop minimal models that admit exact solutions. We prove that the square-root law suppresses the price impact of institutional traders and that price dynamics are diffusive even in the presence of LRC in market-order flow. Furthermore, our minimal models exhibit rich behavior consistent with the ICL and volatility clustering, without introducing any additional mechanisms beyond the square-root law. We thus argue that our exactly solvable framework provides a unified description of price formation under minimal assumptions.

\section{Discrete-time model for price dynamics}\label{sec:discrete-model}
    The main part of this work starts from this section. Let us first study a discrete-time time-series model for price dynamics, by minimally incorporating the price-generating mechanism into the LMF model. Note that a continuous-time version of this model is studied in Sec.~\ref{sec:continuous-time-model}. 

    Because the original LMF model does not have price dynamics, let us specify the price dynamics based on the nonlinear price-impact law $I(Q)=c_0Q^{\delta}$ with general $\delta \in (0,1]$ and $c_0=1$. On the basis of the LMF order-flow dynamics~\eqref{eq:review-LMF-SDE}, the price dynamics is given by the following relation: 
    \begin{box_summary}{Additional assumption for the discrete-time model: price dynamics based on nonlinear price impact}\vspace{-5mm}
        \begin{align}\label{eq:def:discrete-time-model-LMF-impact}
            m(t+\Dt)-m(t) = \eps^{( \ID(t))}(t) \Delta I(Q^{(\ID(t))}(t)),
            \>\>\>m(0)=0,
        \end{align}
    \end{box_summary}\noindent
    where $m(t)$ is the price and $\Delta I(Q):=\left\{ I(Q+1) - I(Q) \right\}$ is the incremental price impact. Note that our model reduces to the square-root-law model if we set $\delta=1/2$ as a special case.

    The state of the discrete-time LMF price-impact dynamics~\eqref{eq:review-LMF-SDE} and~\eqref{eq:def:discrete-time-model-LMF-impact} is completely characterized by a point in the extended phase space 
    \begin{align}
        \mathcal{G}(t):=\left(\eps(t), m(t); \eps^{(1)}(t), R^{(1)}(t), Q^{(1)}(t);\dots;  \eps^{(M)}(t), R^{(M)}(t), Q^{(M)}(t)  \right).
    \end{align}
    Thus, the discrete-time model~\eqref{eq:def:discrete-time-model-LMF-impact} is formulated as a $(3M+2)$-dimensional Markov process.

    \begin{figure*}
        \includegraphics[width=150mm]{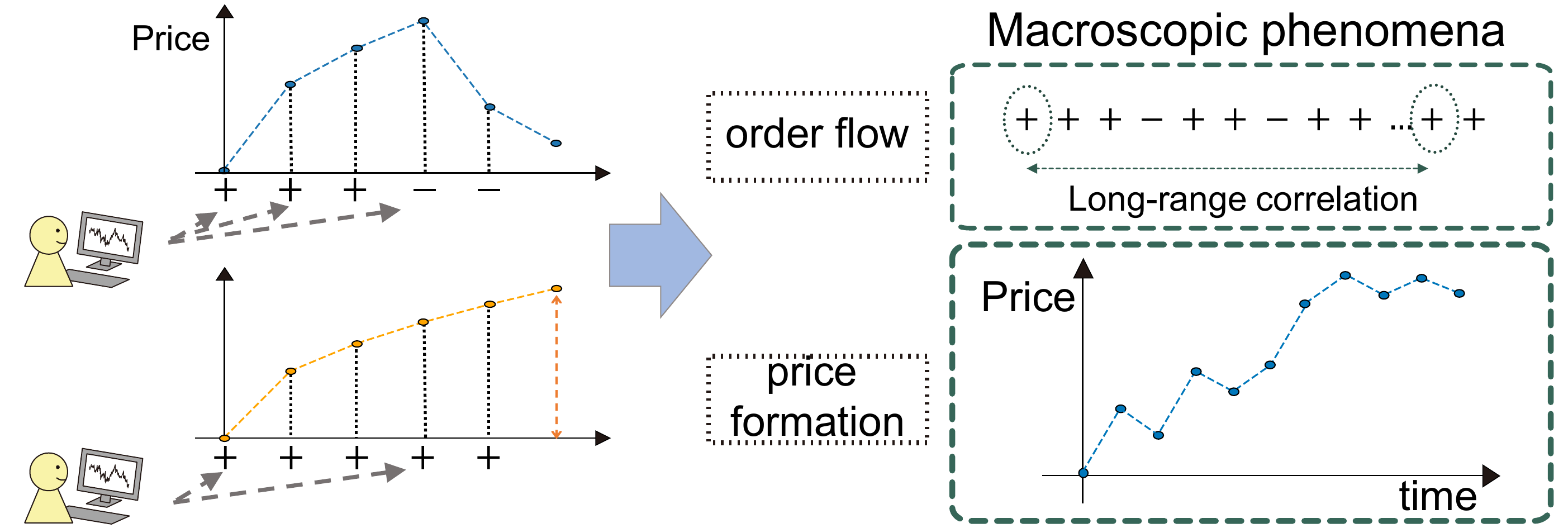}
        \caption{			
            Schematic of the discrete-time LMF price-dynamics model. The system consists of $M$ order-splitting traders. At each time step, one trader is selected with equal probability $\lambda=1/M$ and submits a market order. The price is driven by the sum of the traders' price-impact contributions, according to the additional assumption in Eq.~\eqref{eq:def:discrete-time-model-LMF-impact}. 
        }\label{fig:schematic-discrete}
    \end{figure*}

    In the following subsections, we derive the exact solutions of the discrete-time model. We begin with the single-trader case ($M=1$) to develop our intuition and then generalize to the many-trader case ($M\geq2$). For simplicity, we set $\Dt = 1$ without loss of generality.

    \subsection{Single-trader system ($M=1$)}
        To develop our intuition, let us consider the simplest case with a single trader ($M=1$). Since there is only one trader, trader $i=1$ deterministically submits a market order at each time step. By omitting the superscript (such as $R^{(1)}\to R$), the order-flow dynamics then reduces to
        \begin{subequations}\label{eq:def:discrete-time-LMF-M=1}
            \begin{align}
                R(t+1) &= 
                \begin{cases}
                    R(t) - 1& \mbox{if $R(t)>1$} \\
                    L & \mbox{if $R(t)=1$, $L$ obeys $\psi_{m}(L)$}
                \end{cases}, \\	
                Q(t+1) &= 
                \begin{cases}
                    Q(t) + 1 & \mbox{if $R(t)>1$} \\
                    0 & \mbox{if $R(t)=1$}
                \end{cases}, \\
                    \eps(t+1) &= 
                    \begin{cases}
                    \eps(t) & \mbox{if $R(t)>1$} \\
                        +1 & \mbox{with prob. $1/2$, if $R(t)=1$} \\
                        -1 & \mbox{with prob. $1/2$, if $R(t)=1$}
                \end{cases}.
            \end{align}
            The corresponding price dynamics is given by 
            \begin{align}
                m(t+1)-m(t) = \eps(t) \Delta I(Q(t)),\>\>\>m(0)=0.
            \end{align}
        \end{subequations}
        The price at time $t$ is given by the cumulative sum of the price impacts of individual metaorders, 
        \begin{align}
            \label{eq:discrete-time-reminiscent}
            m(t) = \sum_{u} \eps_u\, I(Q_u), \quad Q_{u} = \min\!\left\{t - t_{u,\ini},\,  t_{u,\fin}-t_{u,\ini}\right\},
        \end{align}
        where $\eps_u$ is the sign of the $u$-th metaorder, and $t_{u,\ini}$ and $t_{u,\fin}$ are its execution start and completion times, respectively.

        \subsubsection{PDF for price dynamics}
            Let us study the PDF of price, denoted by $P_1(x,t)$, where we denote the value of the price $m(t)$ by $x$ for simplicity of notation. After the Fourier-$z$ transformation, the exact expression of the single-trader PDF is given by the following formula:  
            \begin{box_summary}{Discrete-time model for $M=1$: Exact Fourier-$z$ representation of the price-dynamics PDF}
                \vspace{-4mm}
                \begin{align}
                    P_{1}(k,z) := \left(\Fourier_{x\to k} \circ \ZTrans_{t\to z}\right) \left[P_1(x,t)\right]=\Laplace^{-1}_{p\to z}\left[\frac{\Laplace_{z\to p}\left[\Psi_1(k,z)\right]}{1-p\Laplace_{z\to p}\left[\psi_1(k,z)\right]}\right].
                \label{eq:CF-1body}
                \end{align}                    
            \end{box_summary}\noindent

            \paragraph*{Derivation.}            
                Let us introduce the PDF $\eta(x,t)$ as follows~\footnote{In the literature of the L\'evy-walk theory, $\eta(x, t)$ is referred to as the joint probability density; see Ref.~\cite{KlafterReview} and Chapter 8 in Ref.~\cite{KlafterB}.}: the probability that a metaorder is completed during the time interval $[t,t+\dd t)$ and the price is in the range $x\in[x,x+\dd{x})$ is given by $\eta(x, t)\dd x\dd t$. This PDF $\eta(x,t)$ satisfies a recursive relation by construction:
                \begin{align}
                    \eta(x,t) = \int^{\infty}_{-\infty} \dd x_1\sum_{t_{1}=0}^t \eta(x_1,t_1) \psi_{1}(x-x_1, t-t_1) + \delta(x)\delta_{0,t},\quad 
                    \psi_{1}(x,t):= \frac{1}{2}\delta\left(|x|-I(t)\right)\psi_{m}(t), \label{eq:discrete-eta}
                \end{align}
                where $\delta(x)$ is Dirac's delta function, $\delta_{0,t}$ is Kronecker's delta function, $\psi_{1}(x, t)$ is the space-time coupling function, and $I(t)=t^{\delta}$ is the price impact. By using $\eta(x,t)$, the PDF $P_{1}(x,t)$ can be written as
                \begin{align}
                    P_{1}(x,t) =\int^{\infty}_{-\infty} \dd x_{1} \sum_{t_1=0}^{t} \eta(x_{1},t_{1})\Psi_{1}(x-x_1,t-t_1),\>\>\>
                    \Psi_{1}(x,t) := \frac{1}{2}\delta(|x|-I(t))\sum_{t_1=t+1}^{\infty} \psi_{m}(t_1),\label{eq:discrete-PDF}
                \end{align}
                where $\Psi_1(x,t)$ is the space-time survival density\footnote{Here and throughout, we interpret $\delta(|x|-I(t))$ as $\delta(x-I(t))+\delta(x+I(t))$, including at $t=0$.}. By applying the Fourier-$z$ transform to Eq.~\eqref{eq:discrete-PDF} and then applying the Laplace transform with respect to the variable $z$, we obtain Eq.~\eqref{eq:CF-1body}. See Appendix~\ref{app:dm} for a more detailed derivation.

        \subsubsection{Mean-squared displacement}
\begin{figure*}
                \includegraphics[width=150mm]{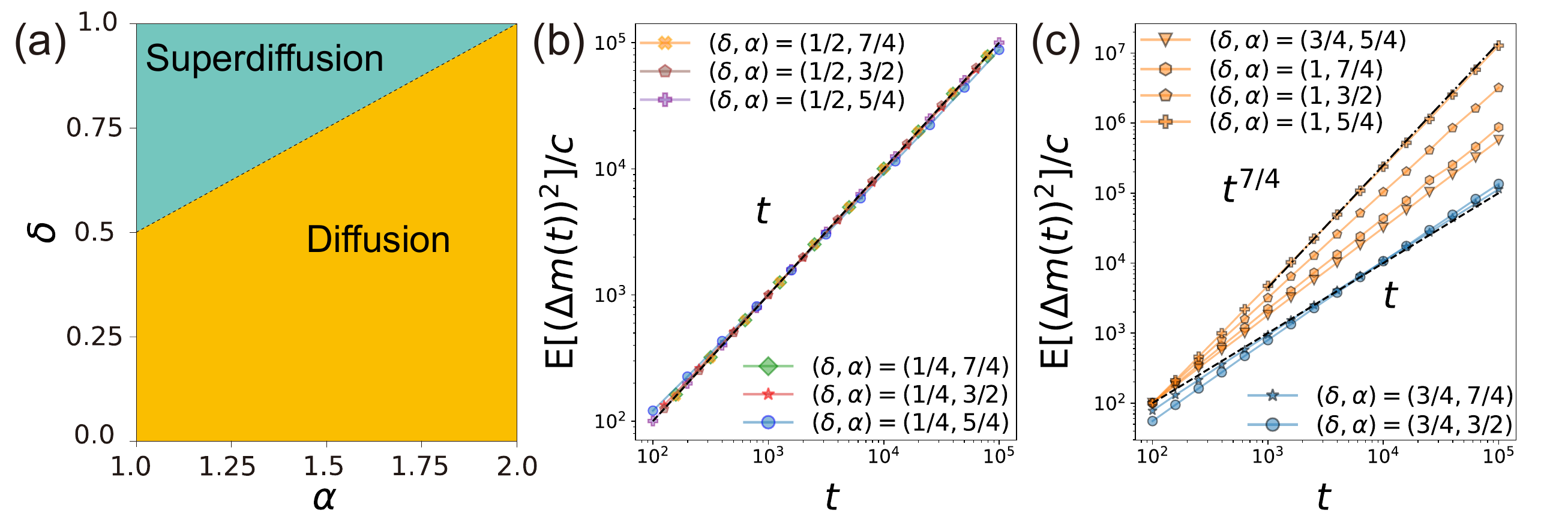}
                \caption{
                Asymptotic behavior of the MSD~\eqref{eq:MSD-discrete-time} for the discrete-time LMF price-impact dynamics~\eqref{eq:discrete-time-reminiscent} for the single-trader case $M=1$. 
                (a)~Phase diagram between superdiffusion ($2\delta>\alpha$) and normal diffusion ($2\delta<\alpha$). The phase boundary is given by $2\delta=\alpha$. Here, $\delta \leq 1/2$ is the necessary and sufficient condition for the MSD to exhibit normal diffusion for any $\alpha>1$. 
                (b)~Numerical MSD for $\delta\in \{0.25,0.5\}$, $\alpha\in \{1.25,1.5,1.75\}$, and $M=1$, showing normal diffusion. 
                (c)~Numerical MSD for $\delta\in \{0.75,1\}$, $\alpha\in \{1.25,1.5,1.75\}$, and $M=1$, showing both superdiffusive cases (orange markers) and normally diffusive cases (blue markers). A logarithmic correction appears at the boundary case $\alpha=2\delta$.
                }\label{fig:MSD-discrete}
            \end{figure*}

            We next report the mean-squared displacement (MSD): 
            \begin{box_summary}{Discrete-time model for $M=1$: Asymptotic formula for the MSD}
                \vspace{-4mm}
                \begin{align}\label{eq:MSD-discrete-time}
                    \rmE[\Delta m(t)]=0,\>\>\>
                    \rmE[\left(\Delta m(t)\right)^2] &\propto 
                    \begin{cases}
                        t                    & \text{if } \alpha > 2\delta \\
                        t^{1+2\delta-\alpha} & \text{if } \alpha < 2\delta
                    \end{cases}
                    \quad \text{for large } t.
                \end{align}                    
            \end{box_summary}\noindent
            For the boundary case, a logarithmic correction appears. See Fig.~\ref{fig:MSD-discrete} for numerical verification. This MSD formula is derived by the following relation:             
            \begin{align}
                \rmE[\Delta m(t)] = \ZTrans_{z\to t}^{-1}\left[i\frac{\partial}{\partial k} P_{1}(k,z)\bigg|_{k=0}\right],\>\>\>
                \rmE[\left(\Delta m(t)\right)^2] = \ZTrans_{z\to t}^{-1}\left[-\frac{\partial^2}{\partial k^2} P_{1}(k,z)\bigg|_{k=0}\right].
            \end{align}           
        
        \subsubsection{Return autocorrelation}
\begin{figure*}
                    \includegraphics[width=180mm]{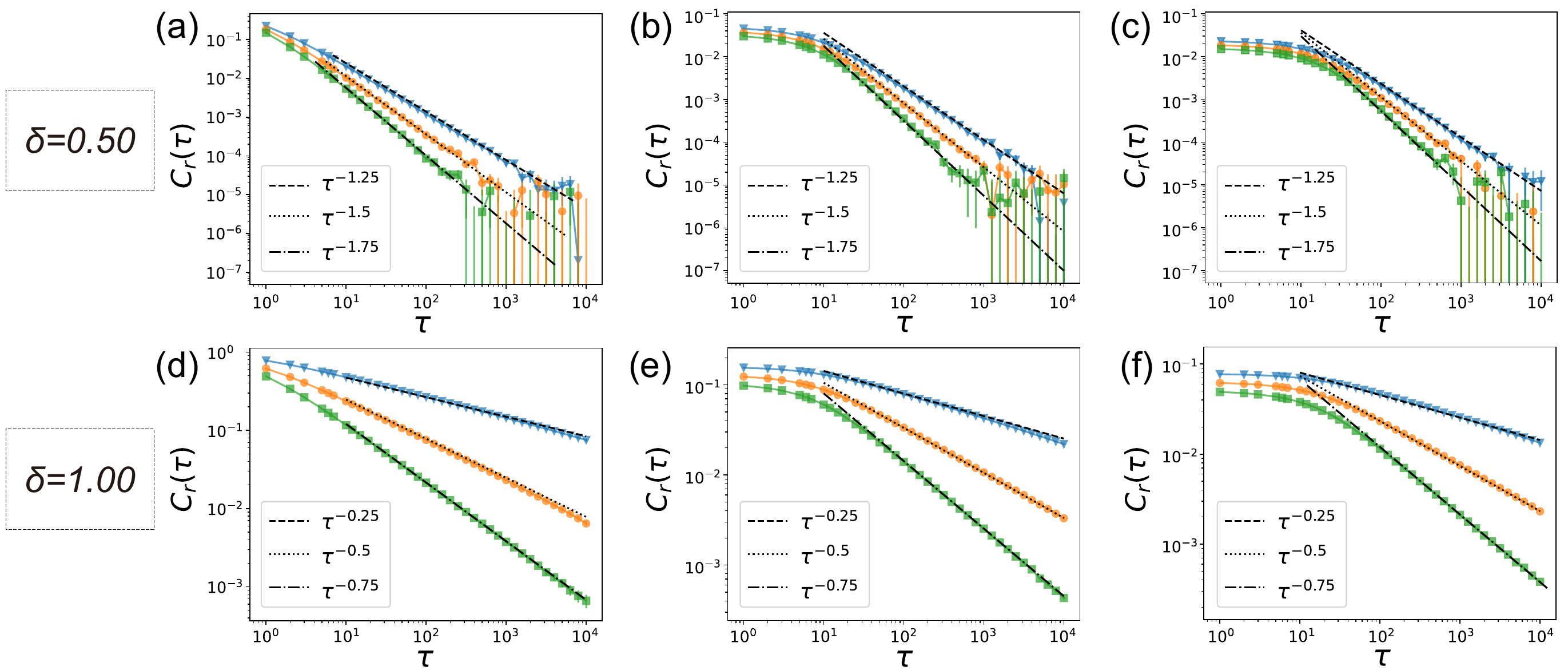}
                    \caption{
                    Numerical results for the price-increment ACF~\eqref{eq:ACF-M=1-asymptotic} and~\eqref{eq:returnACF:discrete-time:Exact}. Panels~(a)--(c) correspond to $\delta=0.5$ and $\alpha\in\{1.25,\,1.5,\,1.75\}$, with (a)~$M=1$, (b)~$M=5$, and (c)~$M=10$. Panels~(d)--(f) correspond to $\delta=1.0$ and $\alpha\in\{1.25,\,1.5,\,1.75\}$, with (d)~$M=1$, (e)~$M=5$, and (f)~$M=10$. All results are consistent with our prediction $C_r(\tau)\propto \tau^{-\theta}$ with $\theta=\alpha-2\delta+1$.
                    }\label{fig:priceacf-discrete}
                \end{figure*}

            We also study the ACF of the one-step return $r(t):=m(t+1)-m(t)$ in the stationary state. The one-step return ACF is given by the following asymptotic formula: 
            \begin{box_summary}{Discrete-time model for $M=1$: Asymptotic formula for the return ACF}\vspace{-5mm}
                \begin{equation}\label{eq:ACF-M=1-asymptotic}
                   C_r(\tau):=\rmE[r(0)r(\tau)]_{\rm{ss}}\propto \tau^{-\theta},\>\>\>\theta =\alpha-2\delta + 1.
                \end{equation}
            \end{box_summary}\noindent
            See Fig.~\ref{fig:priceacf-discrete} for the numerical results.

            \paragraph*{Derivation.}
                Let us introduce an indicator variable $u$, such that $u=1$ if the child orders executed at times $t=0$ and $t=\tau$ belong to the same metaorder; otherwise $u=0$. The one-step return ACF can be rewritten as
                \begin{align}
                    \rmE[r(0)r(\tau)]_{\rm{ss}}
                    = \rmE[\eps(0)\Delta I(Q(0)) \eps(\tau) \Delta I(Q(\tau))]_{\rm{ss}}
                    = \rmE[\mathbb{I}_{u=1}\Delta I(Q(0)) \Delta I(Q(0)+\tau)]_{\rm{ss}}.
                \end{align}
                Approximating the price increment $\Delta I(Q):=(Q+1)^\delta - Q^\delta$ as $\Delta I(Q)\simeq \delta Q^{\delta-1}$ for $Q\gg 1$, we obtain
                \begin{align}
                    \rmE[r(0)r(\tau)]_{\rm{ss}} =& \sum_{Q=0}^{\infty}\sum_{R=1}^{\infty}\mathbb{I}_{R\geq \tau+1} P_{\rm st}(R,Q)\Delta I(Q) \Delta I(Q+\tau)
                         \simeq \delta^2  \sum_{Q=1}^{\infty}\sum_{R=\tau+1}^{\infty} P_{\rm{st}}(R,Q) Q^{\delta-1}(Q+\tau)^{\delta-1} \notag \\
                         \propto& \int_1^\infty Q^{\delta-1}(Q+\tau)^{\delta-\alpha-1}\dd Q 
                         \approx\int_{1/\tau}^\infty y^{\delta-1}(y+1)^{\delta-\alpha-1} \tau^{-\theta}\dd y \propto \tau^{-\theta}, 
                \end{align}
                where the indicator function $\mathbb{I}_{u=1}$ is replaced by $\mathbb{I}_{u=1}=\mathbb{I}_{R\geq \tau+1}$, the joint-stationary distribution $P_{\rm st}(R,Q)$ is given by $P_{\rm{st}}(R,Q)\propto (R+Q)^{-\alpha-1}$ (see Appendix~\ref{app:stationary} for its derivation), and a variable transformation is applied as $Q=\tau y$ on the last line. We then obtain the formula~\eqref{eq:ACF-M=1-asymptotic}.
 
    \subsection{Many-body system}
        We next consider the many-body case with general $M\geq 1$: i.e., the trader set is given by $\SetTrader = \{1,2,\dots,M\}$, and the uniform intensity is given by $\lambda:=1/M$. Let us derive the exact PDF and the MSD for this system.
        
        \subsubsection{PDF for price dynamics}		
            Let us consider the Fourier-$z$ representation of the $M$-body PDF $P_M(k,z):=\left(\Fourier_{x\to k} \circ \ZTrans_{t\to z}\right) \left[P_M(x,t)\right]$. The Fourier-$z$ representation is exactly given by the following formula:
            \begin{box_summary}{Discrete-time model for $M\geq 1$: Exact Fourier-$z$ representation of the price-dynamics PDF}\vspace{-4mm}		    
                    \begin{equation}
                        P_{M}(k,z) = [P_{1}(k,\lambda z)]^M
                        \label{eq:exact-cf-discrete-M}.
                    \end{equation}
                \end{box_summary}

            \paragraph*{Derivation.}
                \begin{figure*}
                    \includegraphics[width=160mm]{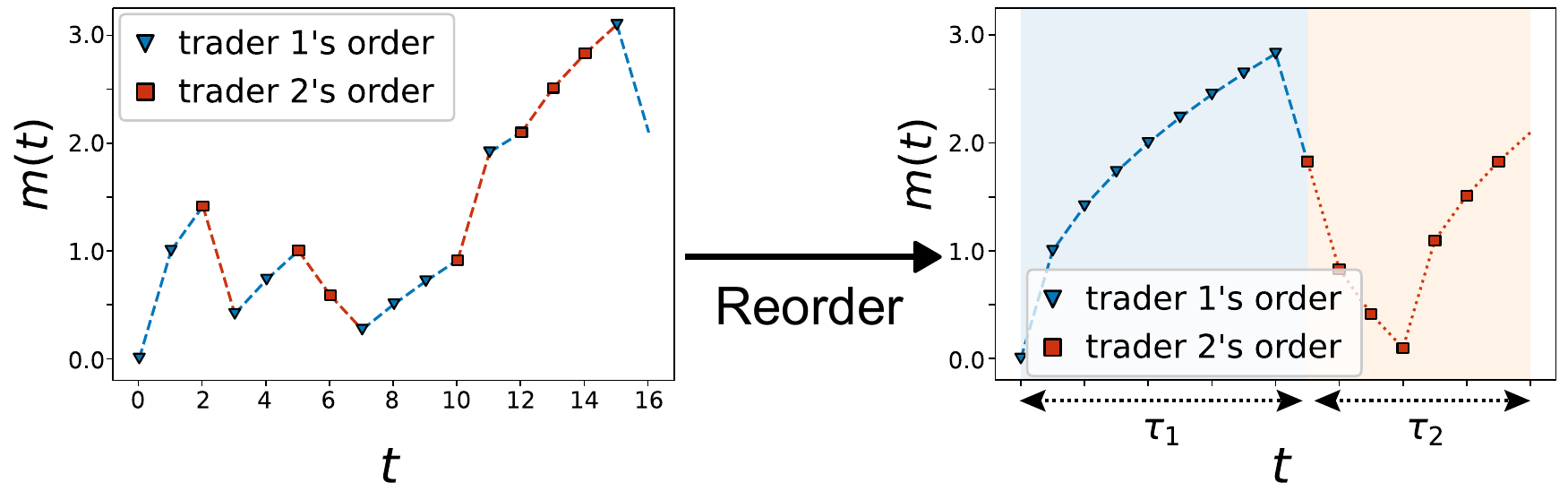}
                    \caption{			
                        Schematic illustration of the reordering procedure used to derive the exact PDF in the two-body system. In the original time series (left), orders from the two traders are randomly interleaved. By reordering the sample path so that trader~$1$ submits all orders during $t \in [0,\tau_1]$ and trader~$2$ submits all orders during $t \in [\tau_1+1,t]$ (right), the price process decomposes into two independent L\'evy-walk contributions.
                    }\label{fig:sol-idea}
                \end{figure*}

                To develop our intuition, let us first consider the two-body system ($\SetTrader = \{1,2\}$). Unlike the single-trader case, the price process $m(t)$ cannot be directly mapped onto the L\'evy walk framework because orders from the two traders are randomly interleaved. However, for a specific sample path in which trader~$1$ submits all orders during $t \in [0, \tau_1]$ and trader~$2$ submits orders during $t \in [\tau_1+1,\, t]$, the price dynamics can be mapped onto the L\'evy walk framework (see Fig.~\ref{fig:sol-idea} for schematic) and decomposed 
                \begin{align}
                    m(t) =
                    \underbrace{\sum_{t_1 \in [0,\tau_1]} \eps^{(1)}(t_1)\, 
        \Delta I(Q^{(1)}(t_{1}))}_{\text{trader 1's contribution}} + \underbrace{        \sum_{t_2 \in [\tau_1+1,\, t]} \eps^{(2)}(t_2)\, \Delta I(Q^{(2)}(t_{2}))}_{\text{trader 2's contribution}}.
                \label{eq:specific-samplepath}
                \end{align}
                The contributions from each trader in~\eqref{eq:specific-samplepath} are independent L\'evy walk processes. The probability of the sample path with identical increments obeys the binomial distribution. Summing over all possible values of $\tau_1$, we obtain the exact PDF as
                \begin{equation}
                    P_{2}(x,t) = \int_{-\infty}^{\infty} \dd{x_1}
                    \sum_{\substack{\tau_1,\tau_2 \geq 0 \\ \tau_1+\tau_2=t}}
                    w(t,\tau_1,\tau_2) P_{1}(x_1,\tau_1)\, P_{1}(x-x_1,\tau_2), \quad 
                    w(t,\tau_1,\tau_2) := \frac{t!}{\tau_1!\tau_2!}\lambda^{t}.
                \end{equation}

                The above calculation can readily be generalized to the $M$-body system. By replacing the binomial distribution $w(t,\tau_1)$ with the multinomial distribution
                \begin{equation}
                    w(t,\tau_1,\dots,\tau_M) := \frac{t!}{\tau_1!\cdots\tau_M!}\lambda^{t}
                \end{equation}
                we obtain 
                \begin{equation}
                P_{M}(x,t) = \int_{-\infty}^{\infty} \dd{x_1}\cdots \int_{-\infty}^{\infty} \dd{x_{M-1}} \sum_{\substack{\tau_1,\dots,\tau_M \geq 0 \\ \tau_1+\cdots+\tau_M=t}} w(t,\tau_1,\dots,\tau_M)\, P_{1}(x-x_{M-1},\tau_{M})\prod_{i=1}^{M-1} P_{1}(x_{i}-x_{i-1},\tau_{i}),\quad x_0:=0.\label{eq:exactPDF}
                \end{equation}
                Applying the Fourier-$z$ transform to Eq.~\eqref{eq:exactPDF}, we obtain Eq.~\eqref{eq:exact-cf-discrete-M}.

        \subsubsection{MSD}
            \begin{figure*}
                \includegraphics[width=180mm]{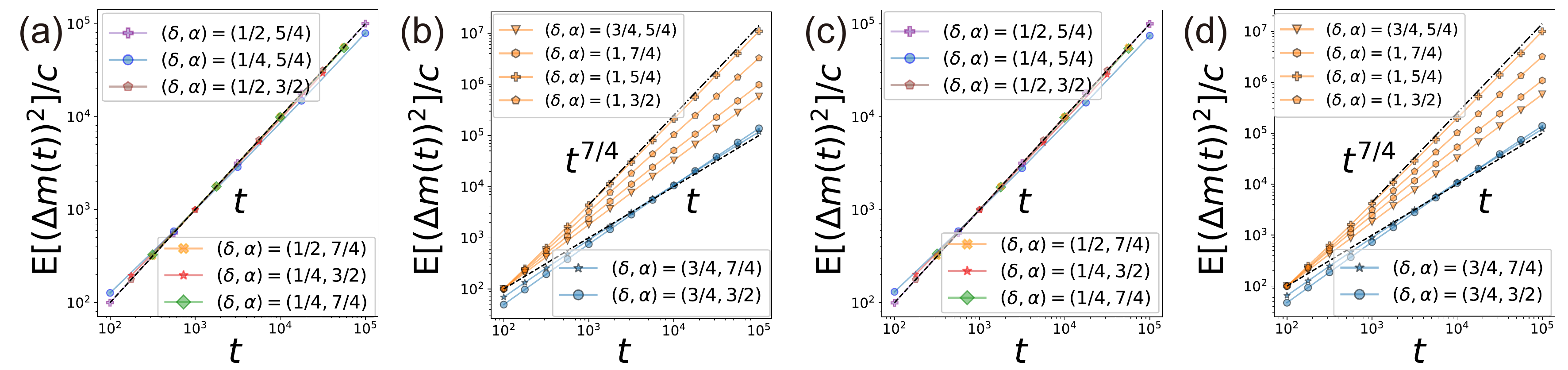}
                \caption{			
                    Numerical result of the discrete-time LMF model~\eqref{eq:def:discrete-time-model-LMF-impact} for the multi-body case~$M>1$. Panels~(a, b) show results for $M=5$, and panels~(c, d) for $M=10$.
                    (a, c)~MSD for $\delta\in \{0.25,0.5\}$, $\alpha\in \{1.25,1.5,1.75\}$, showing normal diffusion.
                    (b, d)~MSD for $\delta\in \{0.75,1\}$, $\alpha\in\{1.25,1.5,1.75\}$, showing the crossover between superdiffusion (orange) and normal diffusion (blue). For the boundary case $2\delta = \alpha$, a logarithmic correction appears.
                }\label{fig:MSD-discrete-many}
            \end{figure*}

            From this expression~\eqref{eq:exact-cf-discrete-M}, we obtain the asymptotic MSD formula identical to that of the single-trader case: 
            \begin{box_summary}{Discrete-time model for $M\geq 1$: Asymptotic formula for the MSD}\vspace{-4mm}
                \begin{align}
                    \label{eq:MSD:discrete-exact}
                    \rmE[\Delta m(t)]=0,\>\>\>
                    \rmE[\left(\Delta m(t)\right)^2]\propto 
                        \begin{cases}
                            t                    & \text{if } \alpha > 2\delta \\
                            t^{1+2\delta-\alpha} & \text{if } \alpha < 2\delta
                        \end{cases}
                        \quad \text{for large } t
                \end{align}
            \end{box_summary}\noindent
            This is the first main result of this work resolving the diffusive price-dynamics paradox. Indeed, it states that $\delta\leq 1/2$ is the necessary and sufficient condition for price dynamics to be diffusive for any $\gamma=\alpha-1\in (0,1)$. Thus, the number of traders is an irrelevant parameter for the long-time diffusive behavior (see Fig.~\ref{fig:MSD-discrete-many} for numerical simulation result).

        \subsubsection{Return ACF}
           The one-step return ACF has a power-law decay even for the many-body case with $M\geq 1$, as an exact asymptotic result: 
            \begin{box_summary}{Discrete-time model for $M\geq 1$: Asymptotic formula for the return ACF}\vspace{-4mm}		
                \begin{equation}\label{eq:returnACF:discrete-time:Exact}
                   C_r(\tau):=\rmE[r(0)r(\tau)]_{\rm{ss}}\propto \tau^{-\theta},\>\>\>\theta =\alpha-2\delta + 1.
                \end{equation}
            \end{box_summary}\noindent
            This is the second main result, suggesting that price returns have short memory when $\alpha>2\delta$; in particular, this condition is satisfied for every $\alpha>1$ if $\delta\leq 1/2$. The integrable correlation is consistent with the applicability of a central limit theorem (CLT) and with the normal-diffusion result. See Fig.~\ref{fig:priceacf-discrete} for numerical simulation results. See Appendix~\ref{app:sec:price-acf} for the detailed derivation. 
            
    \subsection{Implications and remarks}
        We discuss the implications of our findings on the discrete-time model for general $M\geq 1$, particularly from the viewpoint of the diffusive price-dynamics paradox under the LRC. The asymptotic MSD formula~\eqref{eq:MSD:discrete-exact} demonstrates that $\delta\leq 1/2$ is a sufficient condition for the price dynamics to exhibit normal diffusion even in the presence of the LRC. Also, the return ACF~\eqref{eq:returnACF:discrete-time:Exact} exhibits short memory (i.e., $\theta>1$ and $\rmE[r(0)r(\tau)]_{\rm{ss}}$ is integrable) for $\delta\leq 1/2$. The short-memory return ACF is consistent with the applicability of a CLT and with normal Brownian behavior at long times. Thus, the set of the main results~\eqref{eq:MSD:discrete-exact} and~\eqref{eq:returnACF:discrete-time:Exact} shows that the sufficient concavity of nonlinear price impact $I(Q)\propto Q^\delta$ with $\delta\leq 1/2$ is the key to resolve the diffusive price-dynamics paradox under predictable market-order flow. 
        
        Conversely, if the price impact were not sufficiently concave, with $\delta>1/2$, there would exist a threshold $\alpha^*:=2\delta>1$. The price dynamics would be superdiffusive under the LRC for $\alpha\in(1,\alpha^*)$, where the return ACF exhibits long memory, with $\theta\in(0,1)$. In this regime, the usual short-memory argument supporting the CLT no longer applies. Note that the linear price-impact law $\delta=1$ violates the condition for diffusive price dynamics throughout the relevant range $1<\alpha<2$, although classical economic models deduce the linear law.
               
        Here we stress that the square-root law $\delta=1/2$---empirically observed as a universal law across various markets---plays an important role for the diffusive price dynamics: the square-root law suppresses market impacts of large metaorders submitted by institutional investors. While its microscopic origin is still unknown, it is inspiring that the square-root law satisfies the sufficient condition for diffusive price dynamics. 

        In the following sections, we will generalize these discrete-time models to continuous-time models. Our continuous-time models belong to the nonlinear L\'evy-walk models whose analytical properties are well-known. Finally, we find that all the main results~\eqref{eq:MSD:discrete-exact} and~\eqref{eq:returnACF:discrete-time:Exact} can be generalized for all our continuous-time models. 

\section{Continuous-time model for price dynamics}\label{sec:continuous-time-model}
    \begin{figure*}
        \includegraphics[width=180mm]{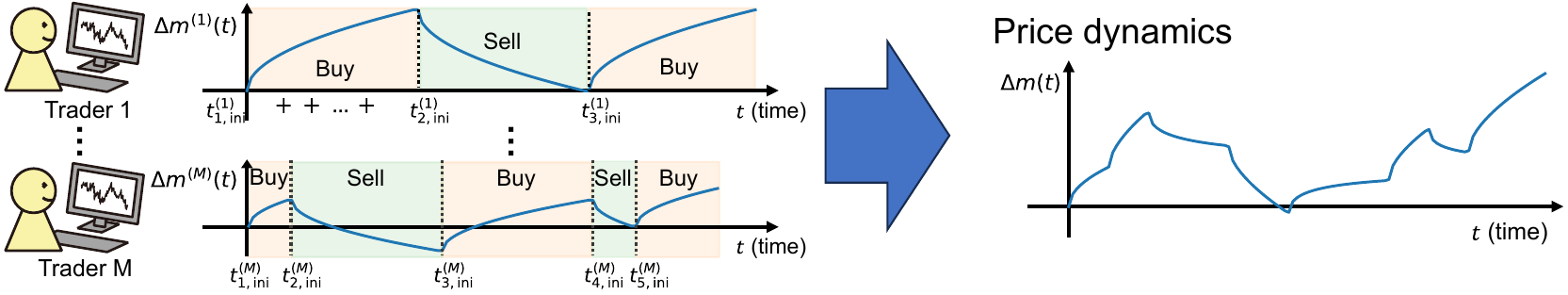}
        \caption{			
            Schematic of the continuous-time model~\eqref{eq:def:continuous-time-LMF-Levy}. Assuming $M$ order-splitting traders with a metaorder size distribution $\psi_{m}(Q) \propto Q^{-\alpha-1}$, the price-impact contribution $\Delta m^{(i)}$ for trader $i$ follows the nonlinear scaling $I(Q) \propto Q^{\delta}$. The total price movement $\Delta m$ results from the independent accumulation of all traders' contributions.
        }\label{fig:schematic-wo-rest}
    \end{figure*}
    In this section, we develop a continuous-time version of the LMF price-dynamics model, corresponding to the discrete-time model~\eqref{eq:review-LMF-SDE} and~\eqref{eq:def:discrete-time-model-LMF-impact}. The advantage of this continuous-time model is that the established theory of nonlinear L\'evy walks is readily applicable and, thus, the formulation is minimal and clearer than the discrete-time model, while keeping all our main findings~\eqref{eq:MSD:discrete-exact} and~\eqref{eq:returnACF:discrete-time:Exact}.  
    
    \subsection{Model}
        As a natural continuous-time extension of the discrete-time model described by Eqs.~\eqref{eq:review-LMF-SDE} and~\eqref{eq:def:discrete-time-model-LMF-impact}, we will consider the following nonlinear L\'evy-walk model (see Fig.~\ref{fig:schematic-wo-rest} for schematic): 
        \begin{box_summary}{Continuous-time LMF price dynamics (based on the nonlinear L\'evy-walk model)}\vspace{-4mm}		    
            \begin{align}\label{eq:def:continuous-time-LMF-Levy}
                m(t):=\sum_{i\in\SetTrader}m^{(i)}(t),\>\>\>m^{(i)}(t):=\sum_{u: t^{(i)}_{u,\ini}\leq t}\eps_{u}^{(i)}I\left(Q^{(i)}_{u}(t)\right) ,\>\>\>I(Q):=Q^{\delta},
            \end{align}
        \end{box_summary}\noindent
        where $t^{(i)}_{u,\ini}$ is the initial time of the $u$-th metaorder submitted by trader $i$, $\eps^{(i)}_u$ is the sign of the $u$-th metaorder, and
        \begin{equation}
            Q^{(i)}_u(t) := \min\bigl\{t - t^{(i)}_{u,\ini}, t^{(i)}_{u,\fin} - t^{(i)}_{u,\ini}\bigr\}.
        \end{equation}
        This continuous-time dynamics is the continuous-time counterpart of the formula~\eqref{eq:discrete-time-reminiscent} in the discrete-time model.  

    \subsection{Exact PDF for the continuous-time model}
        The continuous-time LMF price-dynamics model~\eqref{eq:def:continuous-time-LMF-Levy} can be solved within the established framework of the nonlinear L\'evy-walk theory. Let us study the price-dynamics PDF:
        \begin{box_summary}{Continuous-time LMF price dynamics: Exact price-dynamics PDF formula}\vspace{-4mm}		    
            \begin{align}
                P_M(k,t) = [P_1(k,t)]^M, \quad 
                P_1(k,s) = \frac{\Psi_1(k,s)}{1-\psi_1(k,s)},
                \label{eq:cf-continuous}
            \end{align}
        \end{box_summary}\noindent
        where $P_M(k,t)$ is the Fourier representation of the $M$-body PDF, defined by $P_M(k,t):=\Fourier_{x\to k}\left[P_M(x,t)\right]$ for general $M\geq 1$, and $P_1(k,s)$ is the Fourier-Laplace representation for $M=1$ as $P_1(k,s):=\left(\Fourier_{x\to k}\circ \Laplace_{t\to s}\right)\left[P_1(x,t)\right]$. 
        Also, we introduce the space-time coupling functions
        \begin{equation}
            \psi_1(x,t) := \frac{1}{2}\delta(|x|-I(t))\psi_{m}(t),\quad 
            \Psi_1(x,t):=\frac{1}{2}\delta(|x|-I(t))\int_t^{\infty}\dd t'\psi_m(t')
        \end{equation}
        and their Fourier-Laplace representations
        \begin{equation}
            \psi_1(k,s):= \left(\Fourier_{x\to k}\circ \Laplace_{t\to s}\right)\left[\psi_1(x,t)\right],\quad
            \Psi_1(k,s):= \left(\Fourier_{x\to k}\circ \Laplace_{t\to s}\right)\left[\Psi_1(x,t)\right].
        \end{equation}
        
        \subsubsection*{Derivation}
            As the simplest case, let us study the single-trader case $M=1$. Following the convention of the L\'evy-walk theory (see Ref.~\cite{KlafterReview, KlafterB}), we introduce $\eta(x, t)$ as the probability density\footnote{In other words, $\eta(x, t)\dd t\dd x$ is the probability that a metaorder is completed during the time interval $[t,t+\dd t)$ and then arrives within the range $[x,x+\dd x)$.} that the price reaches $x$ upon completion of a metaorder at time $t$. This probability density $\eta(x,t)$ satisfies a recursive relation:
            \begin{align}
                \eta(x,t) = \int^{\infty}_{-\infty}\dd{x_1} \int^{t}_{0}\dd{t_1} \eta(x_1,t_1)\psi_1(x-x_1,t-t_1) + \delta(x)\delta(t), 
            \end{align}
            which characterizes the price-dynamics PDF $P_1(x,t)$ as
            \begin{align}
                P_{1}(x,t) = \int^{\infty}_{-\infty}\dd{x_1}\int^{t}_{0}\dd{t_1} \eta(x_1,t_1)\Psi_{1}(x-x_1,t-t_1).
            \end{align}
            By applying the Fourier-Laplace transform, we obtain $P_1(k,s):=\left(\Fourier_{x\to k}\circ \Laplace_{t\to s}\right)\left[P_1(x,t)\right]=\Psi_1(k,s)/\{1-\psi_1(k,s)\}$.

            Because the price impact contributions of each trader are independently accumulated, the many-body PDF $P_M(x,t)$ for $M\geq 1$ is given by the convolution relation: 
            \begin{align}
                \label{eq:gen-M-PDF-convolution}
                P_{M}(x,t) = \int^{\infty}_{-\infty}\dd{x_1}\dots\int^{\infty}_{-\infty}\dd{x_{M-1}} P_{1}(x-x_{M-1},t) \prod_{i=1}^{M-1}P_{1}(x_{i}-x_{i-1},t).
            \end{align}
            By applying the Fourier transform, we obtain $P_M(k,t)=[P_1(k,t)]^M$, which is equivalent to Eq.~\eqref{eq:cf-continuous}.

    \subsection{MSD}       
            \begin{figure}
                \centering
                \includegraphics[width=180mm]{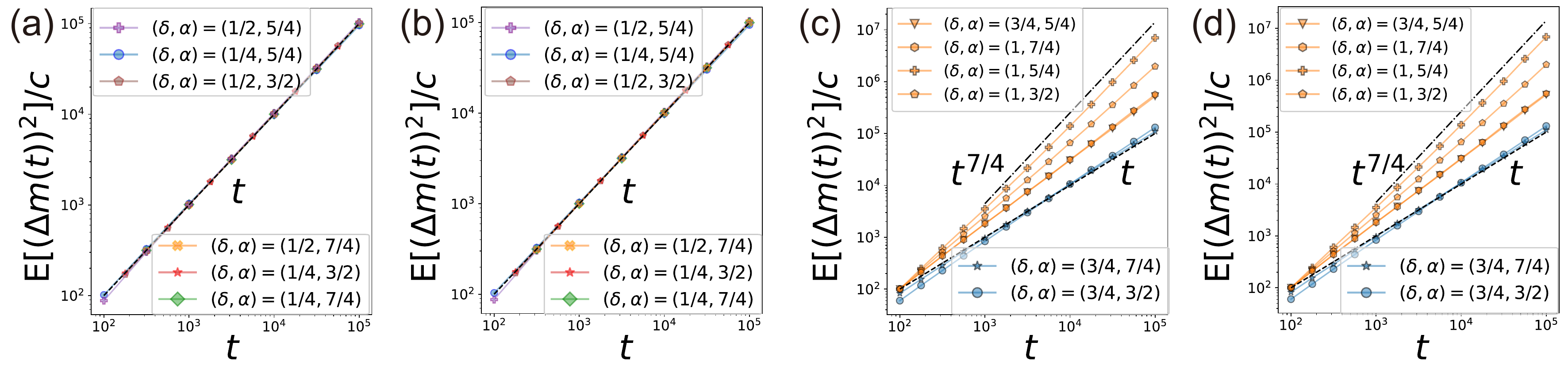}
                \caption{
                    Numerical MSD of the continuous-time LMF price-impact model~\eqref{eq:def:continuous-time-LMF-Levy}.
                    (a, b)~MSD for $\delta\in\{0.25,0.5\}$ and $\alpha\in\{1.25,1.5,1.75\}$, showing normal diffusion. See panel (a) for $M=5$ and (b) for $M=10$.
                    (c, d)~MSD for $\delta\in\{0.75,1.0\}$ and $\alpha\in\{1.25,1.5,1.75\}$, showing the transition between superdiffusion (orange) and normal diffusion (blue). See panel (c) for $M=5$ and (d) for $M=10$. 
                    For the boundary case $2\delta = \alpha$, a logarithmic correction appears.
                }\label{fig:diffusion-wo-rest}
            \end{figure}

            Using the exact formula~\eqref{eq:cf-continuous}, the MSD can be asymptotically evaluated as 
            \begin{box_summary}{Continuous-time LMF price dynamics: Asymptotic MSD formula}\vspace{-4mm}
                \begin{align}\label{eq:continuousLMF-MSD}
                    \rmE[\Delta m(t)]=0,\>\>\>
                    \rmE[\left(\Delta m(t)\right)^2] \propto 
                        \begin{cases}
                            t                    & \text{if } \alpha > 2\delta \\
                            t^{1+2\delta-\alpha} & \text{if } \alpha < 2\delta
                        \end{cases},
                        \quad \text{for large } t.
                \end{align}
            \end{box_summary}\noindent
            See Appendix~\ref{app:cm} for its explicit derivation. This result corresponds to the first main result~\eqref{eq:MSD:discrete-exact} for the discrete-time model~\eqref{eq:def:discrete-time-model-LMF-impact}, showing its robustness even for the continuous-time version. See Fig.~\ref{fig:diffusion-wo-rest} for numerical verification. 

    \subsection{Velocity correlation}
        Let us study the velocity correlation of the price dynamics ($v_M(t):=\dd m(t)/\dd t$), which is the continuous-time counterpart of the one-step return ACF in the discrete-time model. For the single-trader system, the velocity correlation is obtained from the renewal velocity process studied by Ref.~\cite{MeyerBarkaiKantz2017}. In the stationary state, the velocity correlation asymptotically decays as
        \begin{align}
            \rmE\!\left[v_{1}(t)v_{1}(t+\tau)\right]_{\rm ss} \propto \tau^{-\theta},\qquad            \theta = \alpha - 2\delta + 1 . \label{eq:velocity-ACF-single}
        \end{align}
        Since the total velocity in the many-trader system is written as $v_{M}(t)=\sum_{i\in\SetTrader}v^{(i)}(t)$, the corresponding velocity correlation is given by 
        \begin{align}
            \rmE\!\left[v_{M}(t)v_{M}(t+\tau)\right]_{\rm ss} &= 
            \sum_{i,j\in\SetTrader}\rmE\!\left[ v^{(i)}(t)v^{(j)}(t+\tau)\right]_{\rm ss}  =\sum_{i\in\SetTrader}\rmE\!\left[ v^{(i)}(t)v^{(i)}(t+\tau)\right]_{\rm ss} + \sum_{i,j \in\SetTrader, i\neq j}\rmE\!\left[ v^{(i)}(t)v^{(j)}(t+\tau)\right]_{\rm ss} \notag \\
            &\propto M \tau^{-\theta},\qquad
            \theta = \alpha - 2\delta + 1 . \label{eq:velocity-ACF-multi}
        \end{align}
        Here, the cross terms with \(i\neq j\) vanish because different traders are independent and $\rmE[v^{(i)}(t)]_{\rm ss}=0$ owing to the symmetric order signs. This result is consistent with the discrete-time return ACF in Eq.~\eqref{eq:returnACF:discrete-time:Exact}.

\section{Further generalization}\label{sec:further-generalization}
    In this section, we introduce two generalized continuous-time models based on the LMF L\'evy-walk model~\eqref{eq:def:continuous-time-LMF-Levy}. First, we generalize the LMF L\'evy-walk model~\eqref{eq:def:continuous-time-LMF-Levy} by taking into account the resting time between events. We then find that such generalized models are consistent with the ICL and the volatility clustering~\cite{BouchaudText}---two enigmatic empirical laws in the field of market microstructure---although we do not introduce any ad hoc mechanisms to replicate these laws. As the second model, we develop a model with impact decay after metaorder completion. Finally, we find that the diffusive nature of price dynamics is maintained even when impact decay is included, provided that the permanent impact is non-zero. 
  
    \subsection{Generalized model I: metaorder splitting with the resting state}
        For the generalized model I, we incorporate an inter-metaorder resting time into the continuous-time L\'evy-walk model~\eqref{eq:def:continuous-time-LMF-Levy}. In the original LMF model, all traders are assumed to start splitting their new metaorder once their previous metaorder execution is completed. However, it is a more plausible assumption that each trader takes a rest after completion of a metaorder. As the generalized model I, we will take this empirical ingredient into account. 
        \begin{figure*}
            \includegraphics[width=180mm]{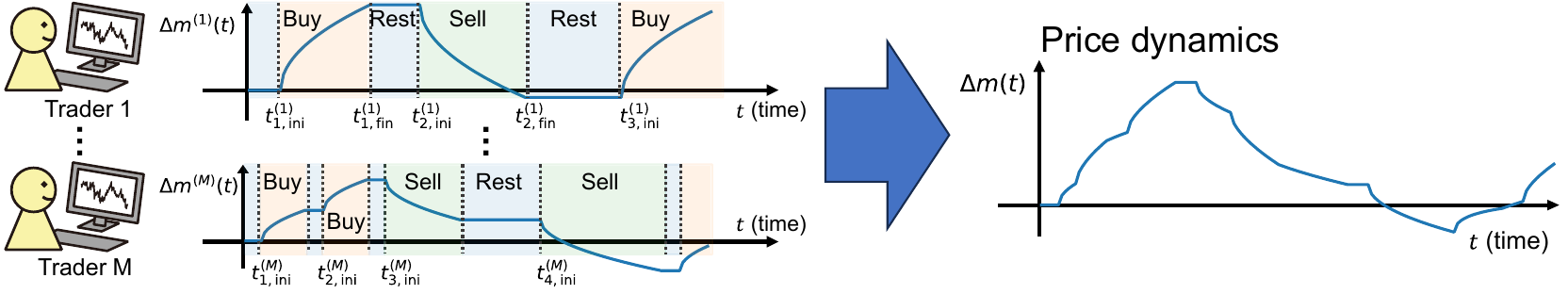}
            \caption{			
                Schematic illustration of the generalized model~I, consisting of $M$ traders. In addition to the base model~\eqref{eq:def:continuous-time-LMF-Levy}, we assume that each trader remains inactive for a random resting time after completing a metaorder execution. The inter-metaorder resting time $\tau$ is drawn from the exponential PDF~\eqref{eq:resting-timePDF-generalized-model-I}. After this resting period, the trader begins executing a new metaorder.
            }\label{fig:Schematic-with-rest}
        \end{figure*}
        \subsubsection{Motivation and other enigmatic laws in finance}
            Let us explain our motivation to generalize the continuous-time L\'evy-walk model~\eqref{eq:def:continuous-time-LMF-Levy}. By taking into account the inter-metaorder resting time, we find that our generalized model I is also consistent with two other enigmatic laws in financial market microstructure: the ICL and the volatility clustering. 

            The ICL is a widely observed empirical law for various assets~\cite{ICLLetter,ICLCompany,ICLIndex}. It states that the price change $\Delta m:=m(t+\Delta t)-m(t)$ obeys a power-law CCDF with an exponent typically near three: 
            \begin{box_summary}{Empirical law: the ICL}\vspace{-4mm}
            \begin{align}
                P_{\geq}(\Delta m,\Delta t):= \int_{\Delta m}^\infty P(\Delta m',\Delta t)\dd (\Delta m') \propto \left(\Delta m\right)^{-\beta}, \quad \beta \approx 3 \quad \mbox{for } \Delta m\to +\infty,
            \end{align}
            \end{box_summary}\noindent
            where $P_{\geq}(\Delta m,\Delta t)$ is the CCDF corresponding to the PDF $P(\Delta m,\Delta t)$ and $\beta$ is the characteristic power-law exponent\footnote{Although $\beta$ is typically observed to be close to three, we do not assume that it is a strictly universal exponent.}. This empirical law is typically observed for time window $\Delta t$ ranging from roughly one minute to several days. The microscopic mechanism behind the ICL has remained a mystery. Note that the price-change PDF is typically symmetric and the ICL is expected to hold even for $\Delta m\to -\infty$.

            In addition, let us explain the volatility clustering. The volatility clustering refers to the long-range correlation of the squared price increments, quantified by the normalized ACF:   
            \begin{box_summary}{Empirical law: the volatility clustering}\vspace{-4mm}
                \begin{align}
                    C_V(\tau):= \frac{\rmE\left[ (\sigma^2(0,t)-\mu_{\sigma^2_t})(\sigma^2(\tau,t+\tau)-\mu_{\sigma^2_t}) \right]}{\rmE[(\sigma^2(0,t)-\mu_{\sigma^2_t})^2]} 
                    \propto \tau^{-\xi},\quad 0<\xi<1,
                \end{align}
            \end{box_summary}\noindent
            where $\mu_{\sigma^2_t} := \rmE[\sigma^2(0,t)]$ and $\sigma^2(t,t+\tau) := \{p(t+\tau) - p(t)\}^2$. Although this law is widely observed for various assets, there is no consensus about its microscopic origin yet. 

            As shown in the subsequent section, we show that these two empirical laws can be replicated by our generalized model I simply by adding the inter-metaorder resting time distribution. 

        \subsubsection{Model}
            Let us assume that each trader remains idle for a time period $\tau$ after completing a metaorder (see Fig.~\ref{fig:Schematic-with-rest} for a schematic), where $\tau$ is drawn from the exponential distribution: 
            \begin{box_summary}{Additional assumption for the generalized model I: inter-metaorder resting time distribution}\vspace{-4mm}
                \begin{align}\label{eq:resting-timePDF-generalized-model-I}
                    \psi_{r}(\tau):= \frac{1}{\tau_{r}}e^{-\tau/\tau_{r}}.
                \end{align}
            \end{box_summary}\noindent
            During the resting period, the trader submits no orders and thus does not contribute to the price dynamics. After the resting period, the trader initiates a new metaorder. We develop the generalized model~I, by incorporating this exponential inter-metaorder resting time distribution into the continuous-time LMF model~\eqref{eq:def:continuous-time-LMF-Levy}. 

            The generalized model~I is an exactly solvable two-state L\'evy-walk model, in which the two states---price impact and rest---are visited cyclically in this order (see Fig.~\ref{fig:Schematic-with-rest} for a schematic). See Appendix~\ref{app:gm1:cf} for the exact PDF of the price dynamics. This model exhibits normal diffusion in the long-time limit for $\alpha>2\delta$, as shown by the MSD formula~\eqref{eq:continuousLMF-MSD}; see Appendix~\ref{app:gm1:msd} for its derivation and numerical verification.

        \subsubsection{Main result 1: the ICL}        
            \begin{figure*}
                \includegraphics[width=150mm]{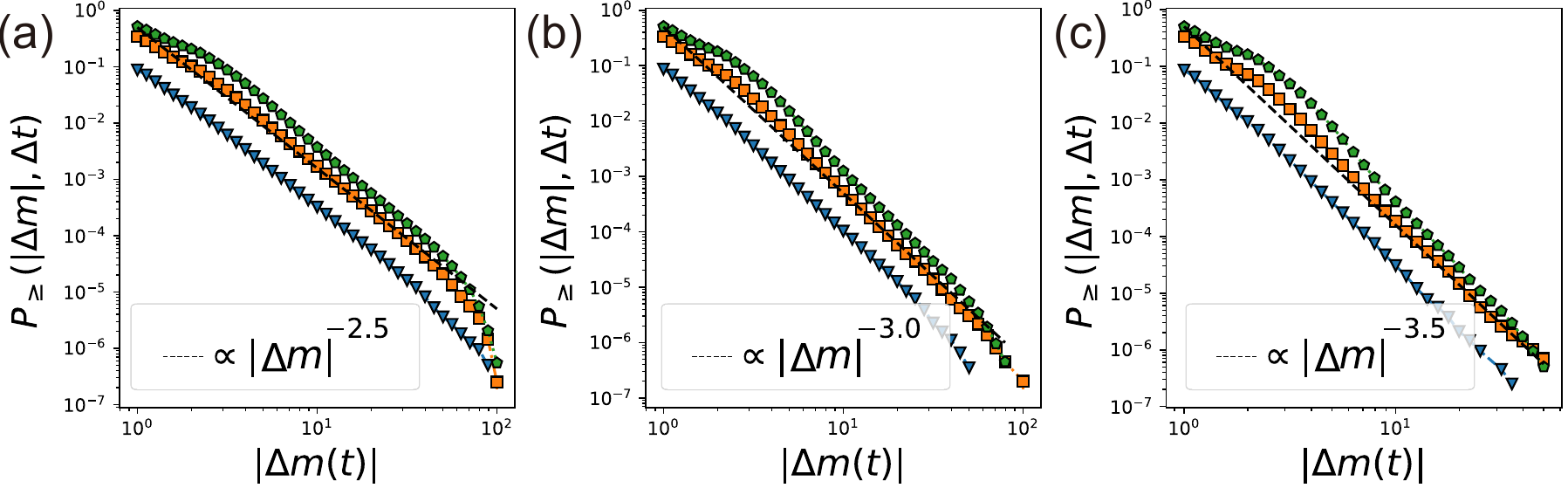}
                \caption{
                    Numerical absolute price-change CCDF of the generalized model~I for $\delta\in\{0.50,1.00\}$, $\alpha\in\{1.25, 1.50, 1.75\}$, and $M\in\{1,5,10\}$. The blue lines represent the CCDF for $M=1$, the orange lines represent the CCDF for $M=5$, the green lines represent the CCDF for $M=10$, and the black lines represent our theoretical lines. The detailed parameters are summarized as follows: (a)~$(\delta,\alpha,\beta,\Delta t,\tau_r)=(0.5, 1.25, 2.5, 10^4, 10^5)$, (b)~$(\delta,\alpha,\beta,\Delta t,\tau_r)=(0.5, 1.5, 3, 10^4, 10^5)$, and (c)~$(\delta,\alpha,\beta,\Delta t,\tau_r)=(0.5, 1.75, 3.5, 10^4, 10^5)$.
                }
                \label{fig:PriceChanges}
            \end{figure*}
            Interestingly, the generalized model~I is consistent with the ICL and the volatility clustering. Indeed, we can analytically derive the ICL by a long resting-time asymptotic limit:  
            \begin{box_summary}{The generalized model I: the ICL for $1\ll \Delta t \ll \tau_r$}\vspace{-4mm}
                \begin{equation}
                    P_{\geq}(\Delta m,\Delta t) \propto (\Delta m)^{-\beta}, \quad \beta:=\frac{\alpha}{\delta}, \quad \mbox{ for $1\ll \Delta t \ll \tau_r$ and $1\ll |\Delta m| \ll (\Delta t)^{\delta}$}.\label{eq:ICL-generalized-model1}
                \end{equation}
            \end{box_summary}\noindent
            See Appendix~\ref{app:gm1:LRTapprox} for details. For $\beta<2$, the power-law tail can be robustly observed for large $t$ due to the generalized CLT. For $\beta > 2$, on the other hand, the power-law tail gradually reduces to Gaussian due to the conventional CLT for very large $t$. Thus, the power-law tail with $\beta>2$ is observed only in a parameter-dependent transient regime as an intermediate asymptotic law. 

            Historically, various microscopic theories have been proposed to explain this stylized fact. The first explanation was based on a traditional economic theory by Gabaix {\it et al.}~\cite{GabaixNature,Gabaix2006}, which relates the tail exponent $\beta$ to the metaorder-size exponent $\alpha$ in a manner similar to our model.\footnote{However, we note that a key prediction of Gabaix's model---namely, that $\delta$ should equal $\alpha-1$---was recently refuted in our previous Letter~\cite{SatoPRL2025}.} Some researchers argue that price dynamics follow a Kesten process driven by trend-following behavior (see the dealer model~\cite{PhysRep2022,Takayasu1992,KzPRL2018}), while others suggest that the self-exciting nature of order flows plays a crucial role (see the nonlinear Hawkes process~\cite{BlancQF2017,KzDiderPRL2021,KzDiderPRR2023}, which essentially belongs to the nonlinear Kesten family). Despite these efforts, no consensus has been reached regarding the microscopic origin of the ICL. Although our model incorporates neither a self-exciting mechanism nor intraday patterns---both of which we consider important for accurately predicting $\beta$---it offers a simple and plausible mechanism for the near-universal observation of $\beta \approx 3$.
            
        \subsubsection{Main result 2: the volatility clustering}
            \begin{figure*}
                    \includegraphics[width=180mm]{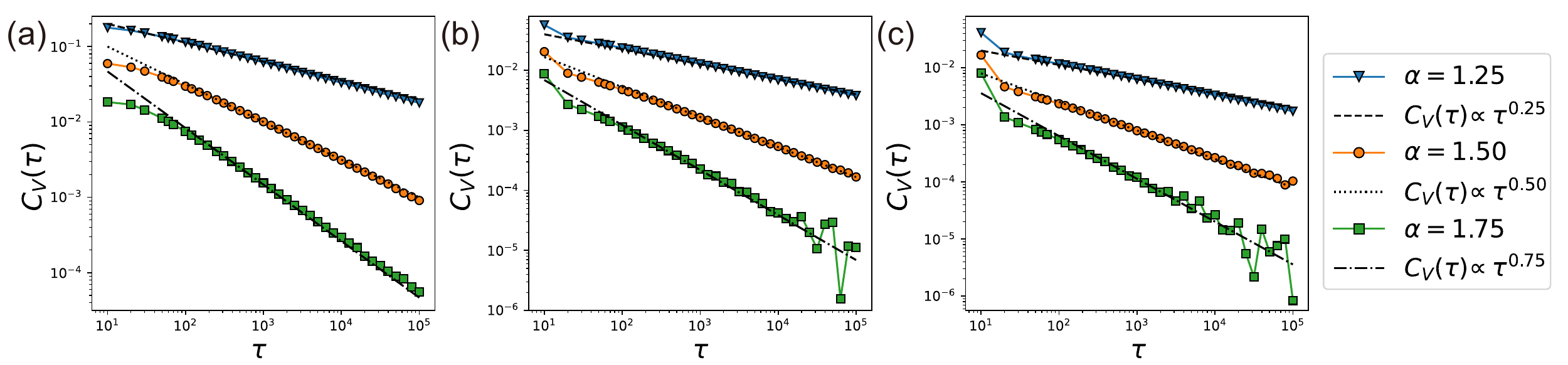}
                    \caption{
                        Volatility ACF in the generalized model~I, numerically showing the long memory $C_{V}(\tau)\propto \tau^{-\xi}$ with $\xi \in (0,1)$. The blue line represents the case with $\alpha=1.25$, the orange line represents the case with $\alpha=1.5$, and the green line represents the case with $\alpha=1.75$. The parameters are summarized as follows: (a)~$(M,\tau_r)=(1,1)$, (b)~$(M,\tau_r)=(5,1)$, and (c)~$(M,\tau_r)=(10,1)$.
                    }\label{fig:volatility}
            \end{figure*}
                
            Furthermore, although we have not obtained a mathematical proof, we numerically find that our model exhibits volatility clustering. Indeed, under the square-root impact law $\delta=1/2$ and by assuming $\alpha \in (1,2)$, we find the following empirical formula (see Fig.~\ref{fig:volatility} for numerical simulation result): 
            \begin{box_summary}{The generalized model I: the volatility clustering as a numerical conjecture}\vspace{-4mm}
                \begin{equation}\label{eq:vol-clustering-generalized-model1}
                    C_V(\tau)\propto \tau^{-\xi},\quad \xi \approx \alpha - 1 \in (0,1).
                \end{equation}
            \end{box_summary}\noindent
            Thus, the volatility in generalized model~I has long memory consistently with empirical observations. 
    
            These two properties~\eqref{eq:ICL-generalized-model1} and \eqref{eq:vol-clustering-generalized-model1} are surprising, given that our model, which relies solely on the plausible assumptions of metaorder splitting and the square-root law, lacks any trivial mechanisms for replicating both ICL and volatility clustering. It is a minimal model, free of artificial memory functions or time-dependent external parameters.
            Note that we have numerically confirmed that the volatility clustering behavior is also robustly observed in the discrete-time, continuous-time, and generalized model II extensions (not shown). 

            \subsection{Generalized model II: resting and impact decay}
        \begin{figure*}
            \includegraphics[width=100mm]{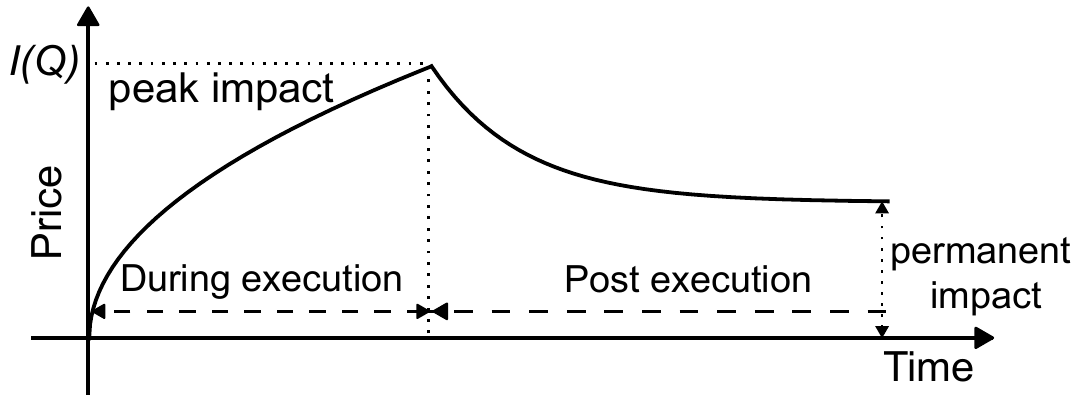}
            \caption{			
                Schematic of the price-impact decay~\eqref{eq:review:impact-decay}. During a buy metaorder execution, the price goes up nonlinearly as a function of the executed volume. After the completion of the metaorder, the impact partially decays on an exponential time scale.
            }\label{fig:schematic-impactdecay}
        \end{figure*}

        \subsubsection{Empirical evidence}
            While we have assumed that the impact does not decay after its peak, let us consider impact decay after its peak by introducing the generalized model II. Some empirical studies report that price impact reaches its peak when the execution is completed and decays after its peak to the permanent impact~\cite{Brokmann2015,Bershova,Bucci2020,Bacry2015}. Although it is not easy to precisely study the functional form of impact decay, one of the most reliable data analyses based on the ANCerno dataset~\cite{Bucci2020} reports that the impact slowly decayed and that the permanent impact was typically one third of the peak impact on average: 
            \begin{equation}
                \label{eq:review:impact-decay}
                G_{\rm decay}(t) = I_{\infty} + (I_{\rm peak}-I_{\infty})e^{-t/\tau_D}, \quad I_{\infty} := \lim_{t\to \infty}I_{\rm decay}(t), \quad \frac{I_{\infty}}{I_{\rm peak}} \approx \frac{1}{3},
            \end{equation}
            where $t$ is the time elapsed after completion of a metaorder, $I_{\rm peak}$ is the maximum impact, $I_{\infty}$ is the permanent impact, and $\tau_{D}\approx 50$ days is the characteristic timescale of decay. 
            
            For the generalized model II, we add this second ingredient to generalized model~I to test whether the diffusivity of price dynamics is maintained even in the presence of impact decay. We confirm that the diffusive nature of price dynamics is maintained if the permanent impact is non-zero $I_{\infty}\neq 0$. 

        \subsubsection{Model}
            We develop the generalized model~II by adding post-metaorder price decay to generalized model~I. In other words, price impact is assumed to decay as the elapsed time $t'$ increases after completion (see Fig.~\ref{fig:schematic-impactdecay} for a schematic), such that
            \begin{box_summary}{Additional assumption for the generalized model II: price impact decay after metaorder completion}\vspace{-4mm}
                \begin{align}
                    \label{eq:impact-decay-generalized-model-II}
                    G(Q,t'):= \rmE[m(t_{\fin}+t')-m(t_{\ini})\mid Q, 0\leq t' \leq \tau] = I(Q)\left\{c_D+(1-c_D)e^{-t'/\tau_{D}}\right\}, \quad 
                    \psi_d(\tau):=\frac{1}{\tau_{d}}e^{-\tau/\tau_d},
                \end{align}
            \end{box_summary}\noindent
            where $t_{\fin}$ is the end time of a previous metaorder execution, $c_D$ is a non-zero constant characterizing the permanent price impact, $\tau_D$ is the characteristic timescale of impact decay, and $\tau$ is the termination time of decay sampled from the PDF $\psi_d(\tau)$ with the characteristic timescale $\tau_d$. We assume that the typical value of $c_D$ is near one third, based on the previous empirical report~\cite{Bucci2020}\footnote{While $c_D$ controls the impact decay, it does not directly represent the permanent impact. Indeed, by averaging over $\tau$, the permanent impact is given by $\rmE[G(Q,\tau)\mid Q] = I(Q)\{c_D + (1-c_D)\tau_D/(\tau_D+\tau_d)\}$. Thus, Generalized model~II always has a non-zero permanent impact in our setup.}. 

            \begin{figure*}
                \includegraphics[width=180mm]{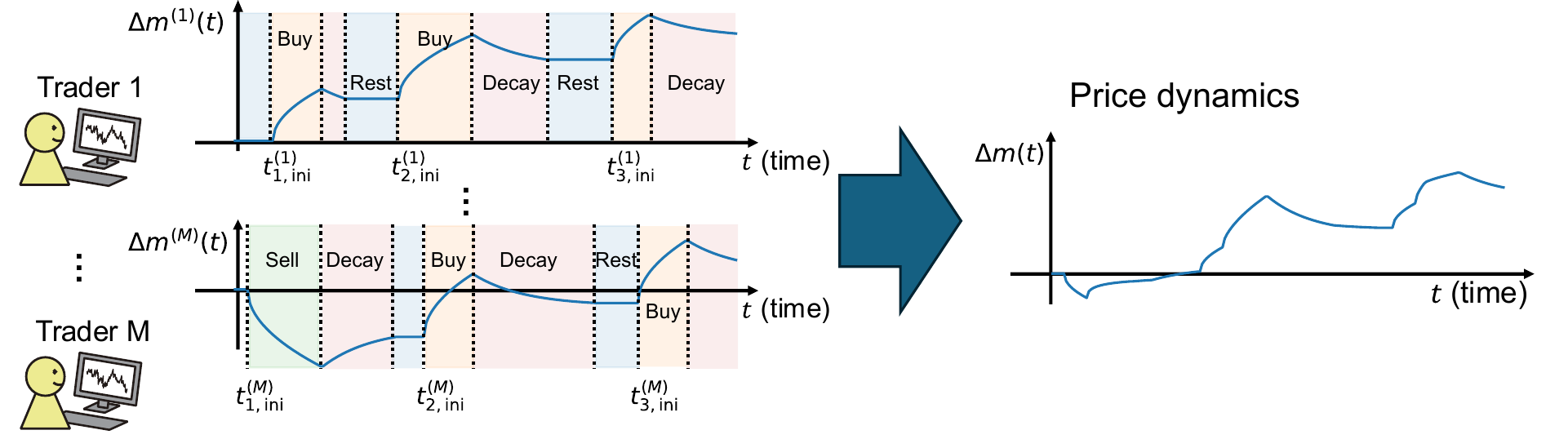}
                \caption{			
                    Schematic illustration of generalized model~II, consisting of $M$ order splitters. Starting from generalized model~I, we further incorporate post-peak impact decay: after reaching its peak, each price impact decays for a finite period according to Eq.~\eqref{eq:impact-decay-generalized-model-II}. 
                }\label{fig:schematic-decay}
            \end{figure*}
        \subsubsection{Exact solutions}
            The generalized model~II is an exactly solvable three-state L\'evy-walk model. The three states---metaorder impact, decay, and rest---are visited cyclically in this order (see Fig.~\ref{fig:schematic-decay} for a schematic). Indeed, we can obtain the characteristic function in a closed form (see Appendix~\ref{app:gm2:cf}). Though the impact decays after metaorder execution, the generalized model~II exhibits essentially the same behavior as the generalized model~I: the MSD formula is identical to Eq.~\eqref{eq:continuousLMF-MSD} and the ICL~\eqref{eq:ICL-generalized-model1} holds likewise. 

        \subsection{Remarks}
            One important assumption of our models is the presence of a nonzero permanent impact component. Under this assumption, our model provides a minimal account of the diffusive price-dynamics paradox in the presence of LRC in market-order flow. It also accounts for other enigmatic empirical laws in financial market microstructure, such as the ICL and volatility clustering, without introducing any ad hoc mechanisms specifically designed to reproduce these laws. Thus, our minimal model offers a simple theory that resolves several paradoxes in finance from a unified viewpoint.
            
            At the same time, we note a potentially debatable point that requires further empirical analysis in the future. Although we regard the assumption of nonzero permanent impact as plausible on the basis of previous evidence~\cite{Bucci2020}, some researchers argue that permanent price impact might not exist once the alpha component associated with informed trading is extracted\footnote{When modeling price dynamics at an effective level, however, our view is that such an alpha contribution should be included in the stochastic dynamics anyway. Indeed, because most traders seek alpha in reality, it does not seem plausible to consider a hypothetical market model in which no traders seek alpha and the permanent component is therefore exactly zero.}. If the permanent impact is exactly zero, the contribution of metaorder impact disappears at long times, and the price dynamics becomes subdiffusive in our description. Therefore, the relevant model class would be essentially different if the permanent component were zero.
            
            Recently, Ref.~\cite{Maitrier2026} proposed a nonlinear propagator model in which the permanent component is exactly zero and the transient impact decays as a power law, while metaorders exhibit long memory. They argue that the long-range correlations of metaorders compensate for the decaying impact and thereby produce diffusive price dynamics even without any permanent component. Their theory also predicts a delicate fine-tuned relationship between metaorder correlations and the slow impact decay. Testing this prediction with microscopic datasets would be an important future task, as it could help confirm or falsify this theoretical scenario and clarify the associated scientific debate.

    \section{Concluding discussion}\label{sec:conclusion}    
        In this paper, we have developed a theoretical framework for price formation that incorporates two key empirical facts: nonlinear price impact and the LRC of buy--sell market-order flow, $\rmE[\eps(t+\tau)\eps(t)]\propto \tau^{-\gamma}$ with $\gamma\in (0,1)$. By extending the LMF model with nonlinear metaorder impact, $I(Q)=c_0 Q^\delta$ with $\delta\in(0,1)$, we derived exact solutions for the PDF and moments of the price in both discrete and continuous time.

        A central finding of our analysis is that, within our model class, the condition $\delta\leq 1/2$ is necessary and sufficient for the price dynamics to exhibit normal diffusion for all $\gamma\in(0,1)$. In particular, this implies that the price dynamics are diffusive under the square-root law, $\delta=1/2$, regardless of the strength of the LRC in the order flow. The implications of the square-root law for each model are summarized in Table~\ref{tab:srl-summary}. This diffusive behavior stands in sharp contrast to that of linear permanent-impact models, in which persistent order flow generically leads to superdiffusion. Our result provides a simple and natural resolution of the apparent paradox between the EMH and the empirically observed LRC of order flow: sufficiently concave price impact restores diffusive price dynamics within our model class.
        
        We also demonstrated that this diffusive property is remarkably robust. The asymptotic scaling of the MSD is independent of the number of traders $M$ and remains unchanged even when inter-metaorder resting periods and post-metaorder impact decay are incorporated, as formulated in generalized models~I and~II. These results suggest that normal diffusion of prices might be a universal consequence of nonlinear price impact combined with order splitting, rather than a fine-tuned property of a specific model.
        
        Furthermore, we showed that our models are consistent with other enigmatic empirical laws in financial market microstructure, such as the ICL and volatility clustering. In particular, the ICL follows from the combination of the power-law metaorder-size distribution and the square-root impact law, while volatility clustering emerges numerically in our models. We stress that we did not introduce any ad hoc mechanisms designed specifically to reproduce these laws. Thus, our model provides a minimal and unified perspective on several empirical regularities in financial market microstructure.

        \begin{table*}
            \begin{ruledtabular}
            \begin{tabular}{lcccc}
                Model & Normal diffusion & Short-memory return ACF & ICL-type price-change tail & Volatility clustering  \\
                \hline
                Discrete-time  model & Yes & Yes & No & Yes \\
                Continuous-time  model & Yes & Yes & No & Yes \\
                Generalized model~I  & Yes & Yes & Yes & Yes\\
                Generalized model~II  & Yes  & Yes & Yes & Yes \\
            \end{tabular}
            \end{ruledtabular}
            \caption{
                Summary of the stylized facts accounted for by each model under the square-root law, $\delta=1/2$.
            }
            \label{tab:srl-summary}
        \end{table*}
        
    \section*{Author contributions}
        YS developed the numerical simulation code and contributed to the analytical calculations for the continuous-time model and generalized models~I and~II. SF contributed to the exact calculations for the single-trader discrete-time system. Both YS and SF contributed to the analytical calculations for the many-body discrete-time system. KK designed and supervised the project and contributed to the analytical calculations, particularly for the ICL. All authors contributed to writing the manuscript and approved its final version.

    \begin{acknowledgements}
		We thank J.-P. Bouchaud and G. Maitrier for fruitful comments. YS was supported by JSPS KAKENHI (Grant No. 24KJ1328). KK was supported by JSPS KAKENHI (Grant Nos. 21H01560, 22H01141, and 25K01450) and the JSPS Core-to-Core Program (Grant No. JPJSCCA20200001).
	\end{acknowledgements}
    
    \appendix

    \section{Analytical solutions to the discrete-time price-dynamics model~\eqref{eq:def:discrete-time-model-LMF-impact}}
        In this Appendix, we describe the detailed analytical calculation for the discrete-time price-dynamics model~\eqref{eq:def:discrete-time-model-LMF-impact}. 
        \subsection{Price-dynamics PDF for the single-trader case $M=1$}\label{app:dm}
            Let us derive the Fourier--$z$ representation of the price-dynamics PDF for the discrete-time model~\eqref{eq:def:discrete-time-LMF-M=1} for the single-trader case $M=1$, by starting from Eq.~\eqref{eq:CF-1body}. Assuming the metaorder-size distribution $\psi_m(t)=t^{-\alpha-1}/\zeta(\alpha+1)$ and the power-law price impact $I(Q) = Q^{\delta}$, we obtain
            \begin{align}
                p\Laplace_{z\to p}\left[\psi_1(k,z)\right]
                &=\frac{1}{\zeta(\alpha+1)}\sum_{t=1}^\infty p^{-t}t^{-\alpha-1}\cos(kt^\delta) = A_0(p)-\frac{k^2}{2}B_0(p) + O(k^4),
            \end{align}
            where we define
            \begin{align}
                A_0(p):=\frac{1}{\zeta(\alpha+1)}\sum_{t=1}^{\infty}p^{-t}t^{-\alpha-1},\quad 
                B_0(p):=\frac{1}{\zeta(\alpha+1)}\sum_{t=1}^{\infty}p^{-t}t^{2\delta-\alpha-1}.
            \end{align}
            $\Laplace_{z\to p}\left[\Psi_1(k,z)\right]$ is obtained as
            \begin{align}
                \Laplace_{z\to p}\left[\Psi_1(k,z)\right]
                \simeq \frac{1}{\zeta(\alpha+1)p}\sum_{t=0}^{\infty}p^{-t}\cos(kt^{\delta})\sum_{t'=t+1}^{\infty}({t'})^{-\alpha-1}
                =A_0'(p)-\frac{k^2}{2}B_0'(p) + O(k^4),
            \end{align}
            where
            \begin{align}
                A_0'(p):=\frac{1}{\zeta(\alpha+1)p}\sum_{t=0}^{\infty}\left(\frac{1}{p}\right)^t\sum_{t'=t+1}^{\infty}({t'})^{-\alpha-1},\quad 
                B_0'(p):=\frac{1}{\zeta(\alpha+1)p}\sum_{t=0}^{\infty}\left(\frac{1}{p}\right)^tt^{2\delta}\sum_{t'=t+1}^{\infty}({t'})^{-\alpha-1}.
            \end{align}
            At $k=0$, this convention gives $A_0'(p)=\{1-A_0(p)\}/(p-1)$ and therefore $\Laplace_{z\to p}[P_1(0,z)]=1/(p-1)$, as required by probability normalization.
            For $\eps\to +0$ with $p=1+\eps$ with non-zero coefficients $B_0^{(0)}, B_0^{(1)}, B_0'^{(0)}$ and $B_0'^{(1)}$, we have asymptotic formulas
            \begin{equation}
            \begin{gathered}
                A_0(p)=1 -\bar{L}\eps + O(\eps^\alpha),\quad 
                B_0(p)=B_0^{(0)} + B_0^{(1)}\eps^{\alpha-2\delta} + \dots,\\
                A_0'(p)=\bar{L} + O(\eps^{\alpha-1}), \quad 
                B_0'(p)=B_0'^{(0)} + B_0'^{(1)}\eps^{\alpha-2\delta-1} + \dots,
            \end{gathered}
            \end{equation}
            by assuming $\alpha \in (1,2)$ and $\delta \in (0,1)$ with $\bar{L}:=\zeta(\alpha)/\zeta(\alpha+1)$, where we use $\sum_{t=1}^\infty e^{-(\ln p)t}t^{2\delta-\alpha-1}\approx \int_0^\infty \dd t e^{-\eps t}t^{2\delta-\alpha-1}=\Gamma(2\delta-\alpha)\eps^{\alpha-2\delta}$ for $2\delta>\alpha$ and $\sum_{t=1}^\infty e^{-(\ln p)t}t^{2\delta}\sum_{t'=t+1}^\infty (t')^{-\alpha-1}\approx \int_0^\infty \dd te^{-\eps t}t^{2\delta}\int_t^\infty \dd t'(t')^{-\alpha-1}\approx \Gamma(2\delta-\alpha+1)\eps^{\alpha-2\delta-1}/\alpha$ for $2\delta>\alpha-1$. Substituting these expressions into the Fourier-$z$ representation of the price-dynamics PDF and expanding to leading order in $k^2$, we obtain
            \begin{align}
                \Laplace_{z\to p}\left[P_1(k, z)\right]
                =\frac{\Laplace_{z\to p}\left[\Psi_{1}(k,z)\right]}{1-p\Laplace_{z\to p}\left[\psi_{1}(k,z)\right]} 
                \simeq (p-1)^{-1}-\frac{k^2}{2}\left\{C_0^{(0)}(p-1)^{-2}+C_0^{(1)}(p-1)^{\alpha-2\delta-2} \right\}.
            \end{align}
            with coefficients $C_0^{(0)}$ and $C_0^{(1)}$. Inverting the Laplace transform, we obtain
            \begin{align}
                P_1(k, z)
                &\simeq e^z\left[1-\frac{k^2}{2}\left\{C_0^{(0)}z+C_0^{(1)}z^{2\delta-\alpha+1}\right\}\right].
            \end{align}
            For a boundary case $2\delta=\alpha$, a logarithmic correction appears for the final MSD formula. At another boundary $2\delta=\alpha-1$, logarithmic corrections appear in subleading terms, which does not change the final MSD formula.
            
        \subsection{Derivation of the joint-stationary distribution for $P_{\rm{st}}(R,Q)$ for the single-trader case $M=1$}\label{app:stationary}
            In this appendix, we derive the joint-stationary distribution for $P_{\rm{st}}(R,Q)$ for the discrete-time price-dynamics model~\eqref{eq:def:discrete-time-LMF-M=1} with $M=1$. Let us derive the joint-PDF based on the master equation approach. The dynamics of $(R,Q)$ is given by 
            \begin{align}
                f(R_{t+1},Q_{t+1})-f(R_t,Q_t) =
                \begin{cases}
                    0 &\mbox{with prob.} 1-\lambda\\
                    f(R_t-1,Q_t+1)-f(R_t,Q_t) &\mbox{with prob. $\lambda$  if $R_t\geq 2$}\\
                    f(L,0)-f(1,Q_t) &\mbox{with prob. $\lambda$ if $R_t=1$}
                \end{cases}
            \end{align}
            where $f(R,Q)$ is an arbitrary smooth function and $L$ is drawn from $\psi_m(L)$. By taking the ensemble avarage, we obtain
            \begin{align}
                \begin{split}
                    \sum_{R,Q}f(R,Q)\Delta_t P(R,Q,t) = 
                    \sum_{R,Q} P(R,Q,t) \lambda \left[\mathbb{I}_{R\geq 2} \left\{f(R-1,Q+1)-f(R,Q)\right\}+\mathbb{I}_{R=1}\sum_{L}\psi_m(L)\left\{f(L,0)-f(1,Q) \right\}\right],
                \end{split}
            \end{align}
            where $\Delta_t P(R,Q,t)=P(R,Q,t+1)-P(R,Q,t)$. By setting $f(R,Q)=\delta_{R,R'}\delta_{Q,Q'}$ for any integers $R', Q'$ and by using a boundary condition $P(R,Q,t)=0$ for $Q<0$, we obtain 
            \begin{align}
                \Delta_t P(R',Q',t) = \lambda \left\{ P(R'+1,Q'-1,t)- P(R',Q',t) +\psi_m(R')\mathbb{I}_{Q'=0}\sum_{Q''}P(1,Q'',t) \right\}.
            \end{align} 
            for any $R'\geq 1$ and $Q'\geq 0$. We then obtain $P_{\rm st}(R,Q)=P_{\rm st}(R+Q,0)$ and $P_{\rm st}(R,0)=\psi_m(R)\sum_Q P_{\rm st}(1,Q)$. Thus, the stationary distribution is given by $P_{\rm st}(R,Q) = (R+Q)^{-\alpha-1}/\zeta(\alpha)$.
        \subsection{Return ACF for the multi-body case $M\geq 1$}\label{app:sec:price-acf}
            We derive the asymptotic one-step return ACF~\eqref{eq:returnACF:discrete-time:Exact} for the multi-body case $M\geq 1$. Let us decompose the one-step return by introducing the trader-selection indicator $\mathbb{I}_{\ID(t)=i}$, such that $r(t) = \sum_{i\in \SetTrader}\mathbb{I}_{\ID(t)=i}\eps^{(i)}(t)\Delta I(Q^{(i)}(t))$, to deduce 
            \begin{align}
                \rmE[r(0)r(\tau)]_{\rm{ss}} =& \sum_{i,j\in \SetTrader}\rmE[\mathbb{I}_{\ID(0)=i}\mathbb{I}_{\ID(\tau)=j}\eps^{(i)}(0)\eps^{(j)}(\tau)\Delta I(Q^{(i)}(0))\Delta I(Q^{(j)}(\tau))] \\
                =& \sum_{i\in \SetTrader}\rmE[\eps^{(i)}(0)\eps^{(i)}(\tau)\Delta I(Q^{(i)}(0))\Delta I(Q^{(i)}(\tau))\mid \ID(0)=\ID(\tau)=i]P(\ID(0)=\ID(\tau)=i),
            \end{align}
            where the non-diagonal terms $i\neq j$ disappear because metaorders issued from different traders are uncorrelated. In addition, we introduce the indicator variable $u_{0,\tau}$ such that $u_{0,\tau}=1$ if the child orders executed at times $0$ and $\tau$ belong to the identical metaorder and $u_{0,\tau}=0$ otherwise. By using $P(\ID(0)=\ID(\tau)=i)=\lambda^2$, the one-step return ACF can be rewritten as
            \begin{equation}  
                \rmE[r(0)r(\tau)]_{\rm{ss}}
                    = \sum_{i\in\SetTrader} \lambda^2 \rmE[u_{0,\tau}\Delta I(Q^{(i)}(0)) \Delta I(Q^{(i)}(\tau))\mid \ID(0)=\ID(\tau)=i]_{\rm{ss}},
            \end{equation}
            because order signs belonging to different metaorders are uncorrelated, while those belonging to the same metaorder are perfectly correlated. Thus, using $M\lambda^2=\lambda$, the one-step autocorrelation can be rewritten as
            \begin{equation}
               \rmE[r(0)r(\tau)]_{\rm{ss}}
                = \lambda\sum_{Q=0}^{\infty}\sum_{R=1}^{\infty}P_{\rm st}(R,Q)\sum_{n=1}^{\min\{\tau,R-1\}}\mathcal{B}_{\tau-1,\lambda}(n-1)\Delta I(Q)\Delta I(Q+n), \quad 
                \mathcal{B}_{\tau,\lambda}(N) = \frac{\tau!}{N!(\tau-N)!}\lambda^N(1-\lambda)^{\tau-N}.
            \end{equation}
            with binomial distribution $\mathcal{B}_{\tau,\lambda}(N)$. Approximating the price increase as $\Delta I(Q) \simeq \delta Q^{\delta-1}$, we obtain 
            \begin{equation}
            \begin{aligned}  
               \rmE[r(0)r(\tau)]_{\rm{ss}}
                =& \lambda\sum_{n=1}^{\tau}\mathcal{B}_{\tau-1,\lambda}(n-1)\sum_{Q=0}^{\infty}\Delta I(Q)\Delta I(Q+n) \sum_{R=n+1}^{\infty}P_{\rm st}(R,Q) \\
                \propto& \sum_{n=1}^{\tau}\mathcal{B}_{\tau-1,\lambda}(n-1)\sum_{Q=1}^{\infty} 
                Q^{\delta-1}(Q+n)^{\delta-\alpha-1} 
                \propto \sum_{n=1}^{\tau}\mathcal{B}_{\tau-1,\lambda}(n-1)n^{2\delta-\alpha-1},
            \end{aligned}
            \end{equation}
            where we use asymptotic formulas
            \begin{align}
                \sum_{R=n+1}^{\infty}P_{\rm st}(R,Q)\propto (Q+n)^{-\alpha}, \quad
                \sum_{Q=1}^\infty Q^{\delta-1}(Q+n)^{\delta-\alpha-1}\approx \int_0^\infty Q^{\delta-1}(Q+n)^{\delta-\alpha-1}\dd Q\propto n^{2\delta-\alpha-1}
            \end{align}
            for large $n$. Since the main contribution of the binomial distribution $\mathcal{B}_{\tau-1,\lambda}(n-1)$ is around $n\approx \lambda \tau$, we approximately obtain $\sum_{n=1}^{\tau}\mathcal{B}_{\tau-1,\lambda}(n-1)n^{2\delta-\alpha-1}\approx (\lambda\tau)^{2\delta-\alpha-1}$.
            This result implies Eq.~\eqref{eq:returnACF:discrete-time:Exact}. 

    \section{Analytical MSD for the continuous-time price-dynamics model~\eqref{eq:def:continuous-time-LMF-Levy}}\label{app:cm}
        In this appendix, we derive the MSD formula~\eqref{eq:continuousLMF-MSD} for the continuous-time LMF nonlinear price-impact model~\eqref{eq:def:continuous-time-LMF-Levy}. As the first step, let us evaluate the characteristic function $P_{1}(k,s) = \Psi_1(k,s)/(1-\psi_1(k,s))$ for the single-trader case $M=1$. The Fourier-Laplace transform of $\psi_1(x,t)$ is given by 
        \begin{align}
            \psi_1(k,s) = \alpha\int^{\infty}_{1}\dd{t}e^{-st} t^{-\alpha-1}\cos(kt^\delta)
            \simeq  \alpha \int^{\infty}_{1}\dd{t}e^{-st}\left\{1-\frac{k^2 t^{2\delta}}{2}\right\}t^{-\alpha-1} 
            \simeq 1-A_{1}s -\frac{k^2}{2}\left\{B_{1} s^{\alpha-2\delta}+C_{1}\right\}
        \end{align}
        with $A_{1}=\alpha/(\alpha-1)$, $B_{1}=\alpha\Gamma(2\delta-\alpha)$, and $C_{1}=\alpha/(\alpha-2\delta)$. $\Psi_{1}(k,s)$ is obtained as 
        \begin{align}
            \Psi_1(k,s) &\simeq \frac{1}{2}\int^{\infty}_{-\infty}\dd{x}\int^{\infty}_{0}\dd{t}e^{-ikx-st} \delta(|x|-t^{\delta})\left\{\mathbb{I}_{t\geq 1}t^{-\alpha} +\mathbb{I}_{t<1}\right\}
            \simeq \int^{\infty}_{0}\dd{t}e^{-st}\left\{1-\frac{k^2 t^{2\delta}}{2}\right\}\left\{\mathbb{I}_{t\geq 1}t^{-\alpha} +\mathbb{I}_{t<1}\right\} + o(k^2) \notag \\
            &\simeq A'_{1} -\frac{k^2}{2}\left\{B'_{1}s^{\alpha-2\delta-1}+C'_{1}\right\}
        \end{align}
        with $A'_{1}=\alpha/(\alpha-1)$, $B'_{1}=\Gamma(1-\alpha+2\delta)$, and $C'_{1}=1/(\alpha-2\delta-1)+1/(1+2\delta)$. We thus obtain
        \begin{align}
            \label{eq:Appendix-gen-model-I:MSD}
            P_1(k,s)=\frac{\Psi_1(k,s)}{1-\psi_1(k,s)}
            = \frac{1}{s}\frac{A'_{1} -\frac{k^2}{2}\left\{B'_{1}s^{\alpha-2\delta-1}+C'_{1}\right\}}{A_{1}+\frac{k^2}{2}\left\{B_{1} s^{\alpha-2\delta-1}+C_{1}s^{-1}\right\}}
            \propto A''_{1} s^{-1} - \frac{k^2}{2}\left\{B_{1}''s^{\alpha-2\delta-2} +C''_{1}s^{-2}\right\}.
        \end{align}
        For large $t$, we obtain 
        \begin{equation}
            \rmE[\Delta m(t)]=\Laplace_{s\to t}^{-1}\left[i\frac{\pd P_1(k,s)}{\pd k}\right]_{k=0}=0,\quad 
            \rmE[(\Delta m(t))^2]=\Laplace_{s\to t}^{-1}\left[-\frac{\pd^2 P_1(k,s)}{\pd k^2}\right]_{k=0} \propto 
                    \begin{cases}
                        t                    & \text{if } \alpha > 2\delta \\
                        t^{1+2\delta-\alpha} & \text{if } \alpha < 2\delta
                    \end{cases}
        \end{equation}
        as the formula for the single-trader case $M=1$. 

        Because the overall price change is the sum of all traders, the MSD for the multi-trader case $M\geq 1$ is $M$ times larger than the one for the single-trader case $M=1$. We thus obtain Eq.~\eqref{eq:continuousLMF-MSD}.         

    \section{Analytical solutions to the generalized model~I}\label{app:gm1}
        In this appendix, we derive moments of the price in the Generalized model~I---the continuous-time model~\eqref{eq:resting-timePDF-generalized-model-I} with a resting state. 
        
        \subsection{Exact PDF}
            The continuous-time model with inter-metaorder resting is exactly solvable by mapping it onto the L\'evy walk framework. We first analyze the single-trader system and then generalize to the $M$-body case.
        
            Let $\eta(x, t)$ be the probability density that the price is at $x$ and a metaorder followed by a resting period is completed at time $t$. By definition, $\eta(x,t)$ satisfies
            \begin{align}
                \eta(x,t) = \int^{\infty}_{-\infty}\dd{x_1}\int^{t}_{0}\dd{t_2} \int^{t_2}_{0}\dd{t_1} 
                \eta(x_1,t_1)\psi_{r}(t_2-t_1)\psi_{1}(x-x_1,t-t_2) + \delta(x)\delta(t).
            \end{align}
            Using $\eta(x,t)$, the joint PDF for single-body system can be written as 
            \begin{equation}
            \begin{gathered}
                P_{1}(x,t) = \int^{t}_{0}\dd{t_1} \eta(x,t_1)\Psi_{r}(t-t_1) + \int^{\infty}_{-\infty}\dd{x_1}\int^{t}_{0}\dd{t_1}\int^{t}_{t_1}\dd{t_2} \eta(x_1,t_1)\psi_{r}(t_2-t_1)\Psi_{1}(x-x_1,t-t_2), \\
                \Psi_r(t):=\int_t^{\infty}\dd t'\psi_r(t')=e^{-t/\tau_r},\quad
                \Psi_{1}(x,t):=\frac{1}{2}\delta(|x|-I(t))\int_t^{\infty}\dd t'\psi_m(t').
            \end{gathered}
            \end{equation}        
            As in the original continuous-time model~\eqref{eq:def:continuous-time-LMF-Levy} without resting, the PDF $P(x,t)$ for the $M$-body system is given by the convolution~\eqref{eq:gen-M-PDF-convolution}.

        \subsection{Characteristic function of the price in the single-body system}\label{app:gm1:cf}
            The characteristic function of the price in the single-trader system is given by 
            \begin{align}
                P_{1}(k,s) = \frac{\Psi_{r}(s)+\psi_{r}(s)\Psi_{1}(k,s)}{1-\psi_{r}(s)\psi_{1}(k,s)}, \quad 
                \psi_{r}(s) = \int^{\infty}_{0}\dd{t}e^{-st} \frac{1}{\tau_{r}}e^{-t/\tau_{r}}
                = \frac{1}{1+s\tau_{r}},
            \end{align}
            where $\psi_{r}(s)$ is the Laplace transform for the resting time distribution $\psi_{r}(t)$. We thus obtain
            \begin{align}
                P_{1}(k,s) 
                = \frac{\Psi_{r}(s)+\psi_{r}(s)\Psi_{1}(k,s)}{1-\psi_{r}(s)\psi_{1}(k,s)}
                = \frac{1}{s}\frac{\tau_{r}+ A_{1}^{'} - \frac{k^2}{2}\left\{B^{'}_{1}s^{\alpha-2\delta-1}+C^{'}_{1}\right\}}{\tau_{r}+A_{1}+\frac{k^2}{2}\left\{B_{1} s^{\alpha-2\delta-1}+C_{1}s^{-1}\right\}}
                \propto A^{''}_{2}s^{-1} - \frac{k^2}{2}\left\{
                    B_{2}^{''}s^{\alpha-2\delta-2}+C_{2}^{''}s^{-2}
                \right\}.
            \end{align}
            Using this asymptotic formula for the characteristic function, we can evaluate the asymptotic behavior of the MSD.

        \subsection{MSD for the many-body system $M\geq 1$}\label{app:gm1:msd}
        \begin{figure}
                \centering
                \includegraphics[width=180mm]{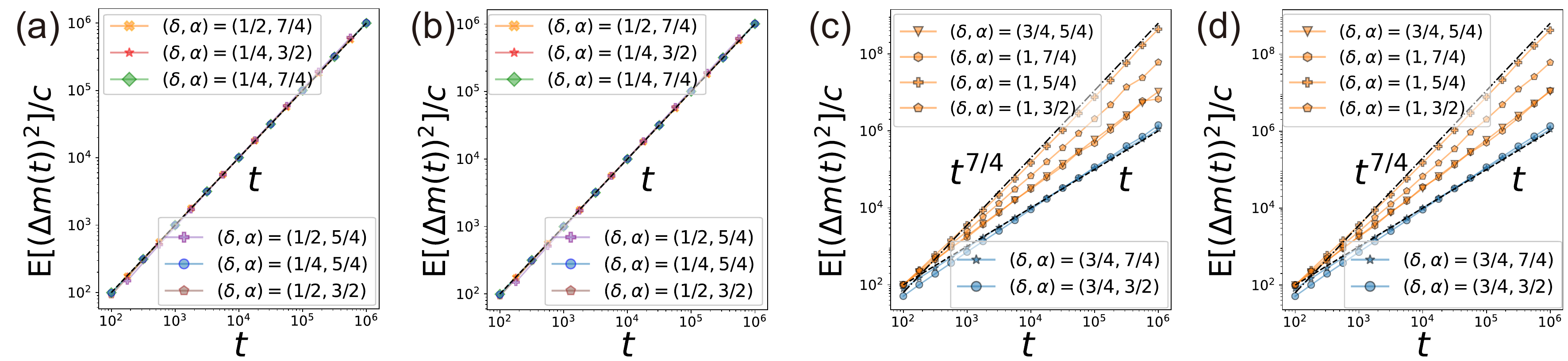}
                \caption{
                    Numerical MSD for the Generalized model~I with $\tau_{r}=100$. (a, b)~Normal diffusion for $\delta\in\{0.25,0.5\}$ and $\alpha\in\{1.25,1.50,1.75\}$. See panel (a) for $M = 5$ and (b) for $M=10$. 
                    (c, d) Crossover between superdiffusion (orange markers) and normal diffusion (blue markers) for $\delta \in \{0.75,1.00\}$ and $\alpha\in\{1.25,1.50,1.75\}$. 
                    See panel (c) for $M=5$ and (d) for $M=10$.
                }
                \label{fig:diffusion-w-rest-multi}
            \end{figure}
            Since the characteristic function for many-body system is given by $P_{M}(k,t) = \prod_{i=1}^M P_{1}(k,t)$. The asymptotic MSD for the multi-trader case $M\geq 1$ is $M$-times larger than the one for the single-trader system (see Fig.~\ref{fig:diffusion-w-rest-multi}). Thus, the asymptotic behavior of the MSD is given by an identical form.

        \subsection{Derivation of the ICL: the long resting-time approximation}\label{app:gm1:LRTapprox}
            Here we derive the asymptotic tail PDF under $\tau_r \gtrsim t$. Given this assumption, the PDF can be approximately obtained 
            \begin{align}
                P(k,s) \approx \Psi_r(s) + \Psi_{1}(k,s)\psi_r(s) + \Psi_r(s)\psi_{1}(k,s)\psi_r(s) + O(\tau_r^{-2}),
            \end{align}
            which implies that
            \begin{align}
                P(x,t) \approx& \Psi_r(t)\delta(x) + \int_0^t \dd t_1 \psi_r(t_1)\Psi_{1}(x,t-t_1) + \int_0^t \dd t_2 \int_{0}^{t_2} \dd t_1 \psi_r(t_1)\psi_{1}(x,t_2-t_1)\Psi_r(t-t_2) + O(\tau^{-2}_r)\\
                =& \delta(x)e^{-t/\tau_r} +  \frac{1}{2 \tau_r \delta} e^{-\frac{t - |x|^{1/\delta}}{\tau_r}} \left( |x|^{-1-\frac{(\alpha - 1)}{\delta}} \mathbb{I}_{|x|\geq 1} \mathbb{I}_{t^\delta \geq |x|}+ |x|^{-1+\frac{1}{\delta}} \mathbb{I}_{1>|x|} \right) \notag \\
                &+ \frac{\alpha }{2\tau_r \delta}e^{-\frac{t-|x|^{1/\delta}}{\tau_r}}\left(t|x|^{-1-\frac{\alpha}{\delta}}-|x|^{-1-\frac{\alpha-1}{\delta}}\right)\mathbb{I}_{|x|\geq 1}\mathbb{I}_{t^\delta \geq |x|} + O(\tau^{-2}_r).
            \end{align}
            On the right-hand side, the first term corresponds to no metaorder executions within the interval $[0,t)$. The second term corresponds to a single metaorder execution ongoing at time $t$. The third term corresponds to a completed metaorder execution, with no subsequent execution, within $[0,t)$. This formula implies an asymptotic power-law form until the cutoff~\eqref{eq:ICL-generalized-model1}.

    \section{Analytical solutions to the generalized model~II}\label{app:gm2}
        In this appendix, we derive analytical solutions to the generalized model~II, which has impact decay~\eqref{eq:impact-decay-generalized-model-II} and a resting state.

        \subsection{Exact PDF}
            Let $\eta(x, t)$ be the joint PDF that the price is at $x$ at time $t$, where $t$ marks the completion of a full cycle consisting of resting, metaorder execution, and impact relaxation. The joint PDF $\eta(x,t)$ satisfies
            \begin{align}
                \eta(x,t) = \int^{\infty}_{-\infty}\dd{x_1}\int^{t}_{0}\dd{t_1}\int^{t}_{t_1}\dd{t_2} \int^{t}_{t_2}\dd{t_3} 
                \eta(x_1,t_1)\psi_{r}(t_2-t_1)\psi_{2}(x-x_1,t_3-t_2,t-t_3) + \delta(x)\delta(t),
            \end{align}
            where $\psi_{2}(x,t_1,t_2)$ is the space-time coupling function characterising the metaorder execution and subsequent impact decay, defined by
            \begin{align}
                \psi_{2}(x,t_1,t_2):= \frac{1}{2}\psi_{m}(t_1)\psi_{d}(t_2)\delta\left(|x|- I(t_1) \left\{c_D + (1-c_D)e^{-t_2/\tau_{D}}\right\}\right).
            \end{align}
            Using $\eta(x,t)$, the joint PDF for single-body system can be written as
            \begin{align}
                \begin{split}
                    P_{1}(x,t) = 
                    &\int^{t}_{0}\dd{t_1} \eta(x,t_1)\Psi_{r}(t-t_1) 
                    + \int^{\infty}_{-\infty}\dd{x_1}\int^{t}_{0}\dd{t_2}\int^{t_2}_{0}\dd{t_1} \eta(x_1,t_1)\psi_{r}(t_2-t_1)\Psi_{1}(x-x_1,t-t_2)\\
                    &+ \int^{\infty}_{-\infty}\dd{x_1}\int^{t}_{0}\dd{t_3}\int^{t_3}_{0}\dd{t_2}\int^{t_2}_{0}\dd{t_1} \eta(x_1,t_1)\psi_{r}(t_2-t_1)\Psi_{2}(x-x_1,t_3-t_2,t-t_3),
                \end{split}
            \end{align}
            where $\Psi_{2}(x,t_1,t_2)$ denotes 
            \begin{align}
                \Psi_{2}(x,t_1,t_2) := \frac{1}{2}\delta\left(|x|-I(t_1)\left\{c_D+(1-c_D)e^{-t_2/\tau_D}\right\}\right)\psi_{m}(t_1)\int^{\infty}_{t_2} \dd t'_2\psi_{d}(t'_2).
            \end{align}
            The PDF $P(x,t)$ for the $M$-body system is given by the convolution~\eqref{eq:gen-M-PDF-convolution}. 

        \subsection{Characteristic function}\label{app:gm2:cf}
            The price-dynamics characteristic function for the multi-body case with $M\geq 1$ is given by $P_{M}(k,t) = \left[P_1(k,t)\right]^M$. Thus, it is sufficient to evaluate the characteristic function $P_1(k,s)$ for the case of a single-trader $M=1$. The characteristic function for the single-trader system is given by 
            \begin{align}
                P_{1}(k,s) = \frac{\Psi_{r}(s)+\psi_{r}(s)\Psi_{1}(k,s)+\psi_{r}(s)}\Psi_{2}(k,s,s){1-\psi_{r}(s)\psi_{2}(k,s,s)}.
            \end{align}

\begin{figure*}
                \includegraphics[width=180mm]{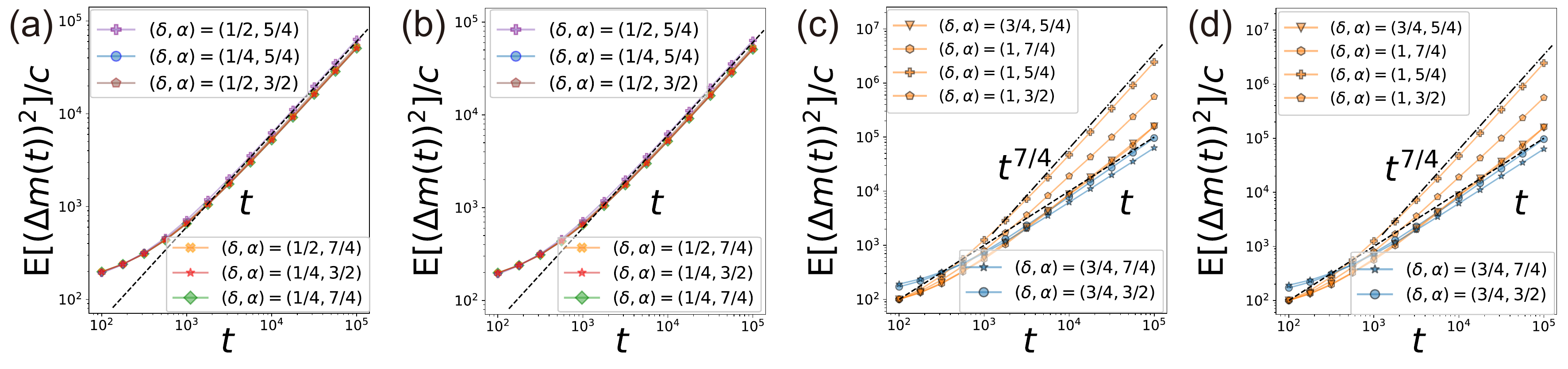}
                \caption{			
                    Numerical MSD for the generalized model~II with $\tau_{r}=100, \tau_{D}=100,  \tau_{d}=100$. 
                    (a, b) Normal diffusion for $\delta\in\{0.25,0.5\}$ and $\alpha\in\{1.25,1.5,1.75\}$. See panel (a) for $M=5$ and (b) for $M=10$. 
                    (c, d) Crossover between superdiffusion (orange markers) and normal diffusion (blue markers) for $\delta \in \{0.75,1.00\}$ and $\alpha\in\{1.25,1.5,1.75\}$. See panel (c) for $M = 5$ and (d) for $M=10$.
                }\label{fig:MSD-trader-5-10}
            \end{figure*}

            Assuming the metaorder-size distribution obeys $\psi_{m}(t)=\alpha t^{-\alpha-1}\mathbb{I}_{t\geq 1}$ and the price impact obeys power-law $I(Q)=Q^{\delta}$, we obtain
            \begin{align}
                \psi_{2}(k,s,s) 
                &= \frac{\alpha}{2\tau_{d}}\int^{\infty}_{-\infty}\dd{x}\int^{\infty}_{1}\dd{t_1}\int^{\infty}_{0}\dd{t_2} e^{-st_1-st_2-ikx}\delta\left( |x|-t_{1}^{\delta}\left(c_{D}+(1-c_{D})e^{-t_2/\tau_{D}}\right)\right) t_1^{-\alpha-1} e^{-t_2/\tau_{d}}\\
                &\simeq \frac{\alpha}{\tau_{d}}\int^{\infty}_{1}\dd{t_1}\int^{\infty}_{0}\dd{t_2} 
                e^{-st_1-st_2} \left( 1 - \frac{t_{1}^{2\delta}k^2}{2} 
                \left(c_{D}+(1-c_{D})e^{-t_2/\tau_{D}}\right)^2\right) t_1^{-\alpha-1} e^{-t_2/\tau_{d}} + o(k^2)\\
                &\simeq 1 -
                A_{3}s
                -\frac{k^2}{2}\left\{B_{3} + C_{3}s^{\alpha-2\delta}\right\},
            \end{align}
            where 
            \begin{align}
            A_{3}=\frac{\alpha}{\alpha-1}+\tau_{d},\>\>\>
            B_{3}=\frac{\alpha(2c_{D}^2\tau_{d}^2+2c_D\tau_d\tau_D+\tau_D(\tau_d+\tau_D))}{(\alpha-2\delta)(\tau_{d}+\tau_{D})(\tau_{D}+2\tau_{d})},\>\>\>
            C_{3}=\frac{\alpha(2c_D^2\tau_d^2+2c_D\tau_d\tau_D+\tau_D(\tau_d+\tau_D))\Gamma(2\delta-\alpha)}{(\tau_d+\tau_{D})(2\tau_{d}+\tau_{D})}        \end{align}
            and $\Psi_{2}(k,s,s) =\tau_{d} \psi_{2}(k,s,s)$. We thus obtain
            \begin{align}
                P_{1}(k,s) 
                &= \frac{1}{s}\frac{
                    \tau_{r}
                    + \left\{A_{1}^{'} -\frac{k^2}{2}\left\{B^{'}_{1}s^{\alpha-2\delta-1}+C^{'}_{1}\right\}\right\} 
                    + \tau_{d}\left\{ 1 -
                A_{3}s
                -\frac{k^2}{2}\left\{B_{3} + C_{3}s^{\alpha-2\delta}\right\}\right\}
                    }{\tau_{r}+A_{3}+\frac{k^2}{2}\left\{B_{3}s^{-1} + C_{3}s^{\alpha-2\delta-1}\right\}}\\
                &\propto A^{''}_{3}s^{-1} - \frac{k^2}{2}\left\{
                        B_{3}^{''}s^{\alpha-2\delta-2}+C_{3}^{''}s^{-2}
                    \right\}.
                    \label{eq:appendix-gen-model-II-PDFScaling}
            \end{align}    
            Based on a discussion similar to~\eqref{eq:Appendix-gen-model-I:MSD}, we obtain the asymptotic MSD scaling~\eqref{eq:continuousLMF-MSD} for general $M\geq 1$ from Eq.~\eqref{eq:appendix-gen-model-II-PDFScaling}, even for the generalized model~II~\eqref{eq:impact-decay-generalized-model-II} (see Fig.~\ref{fig:MSD-trader-5-10} for numerical verification).
           
\end{document}